%% file: main.tex
\documentclass[acmsmall,nonacm]{acmart}
\usepackage{makecell}
\usepackage{ltxtable}
\usepackage{booktabs}
\usepackage{array} 
\usepackage{longtable}
\usepackage{tabularx}
\usepackage{multirow}
\usepackage{caption} 
\usepackage[table]{xcolor}
\usepackage{subcaption}
\usepackage{placeins}
\usepackage{threeparttable}
\newcolumntype{L}[1]{>{\raggedright\arraybackslash}p{#1}}
\newcolumntype{C}[1]{>{\centering\arraybackslash}p{#1}}
\newcolumntype{R}[1]{>{\raggedleft\arraybackslash}p{#1}}

\acmJournal{POMACS}

\title[Multiple Sides of 36 Coins: Measuring Peer-to-Peer Infrastructure Across Cryptocurrencies]{Multiple Sides of 36 Coins: \\ Measuring Peer-to-Peer Infrastructure Across Cryptocurrencies}

\author{Lucianna Kiffer}
\orcid{https://orcid.org/0000-0003-2022-7993}
\affiliation{
  \institution{IMDEA Networks}
  \city{Madrid}
  \country{Spain}
}
\email{lucianna.kiffer@networks.imdea.org}

\author{Lioba Heimbach}
\orcid{https://orcid.org/0000-0002-8258-1712}
\affiliation{
  \institution{Category Labs}
  \city{Zurich}
  \country{Switzerland}
}
\email{lheimbach@category.xyz}
\authornote{This work was partially carried out while the authors were at ETH Zurich.}

\author{Dennis Trautwein}
\orcid{https://orcid.org/0000-0002-8567-2353}
\affiliation{
  \institution{University of Göttingen}
  \city{Göttingen}
  \country{Germany}
}
\email{research@dtrautwein.eu}

\author{Yann Vonlanthen}
\orcid{https://orcid.org/0000-0001-5736-8197}
\affiliation{
  \institution{Ethereum Foundation}
  \city{Zurich}
  \country{Switzerland}
}
\email{yann.vonlanthen@ethereum.org}
\authornotemark[1]

\author{Oliver Gasser}
\orcid{https://orcid.org/0000-0002-3425-9331}
\affiliation{
  \institution{IPinfo}
  \city{Seattle}
  \country{United States}
}
\email{oliver@ipinfo.io}

\date{}

\begin{document}

\begin{abstract}

Blockchain technologies underpin an expanding ecosystem of decentralized applications, financial systems, and infrastructure. However, the fundamental networking layer that sustains these systems, the peer-to-peer (P2P) layer, of all but the top few ecosystems remains largely opaque. In this paper, we present the first longitudinal, cross-network measurement study of 36 public blockchain networks. Over 9 months (since late 2024), we deployed 15 active crawlers, sourced data from two additional community crawlers, and conducted hourly connectivity probes (e.g., pings and protocol-level handshakes) to observe the evolving state of these networks. Furthermore, by leveraging Ethereum’s discovery protocols, we inferred metadata for an additional 19 auxiliary networks that utilize the Ethereum peer discovery protocol. 
We also explored Internet‑wide scans, which only require probing each protocol’s default ports with a simple, network‑specific payload. This approach allows us to rapidly identify responsive peers across the entire address space without having to implement custom discovery and handshake logic for every blockchain. We validated this method on Bitcoin and similar networks with known ground truth, then applied it to Cardano, which we could not crawl directly.

Our study uncovers dramatic variation in network size from under 10 to more than 10,000 active nodes. We quantify trends in IPv4 versus IPv6 usage, analyze autonomous systems and geographic concentration, and characterize churn, diurnal behavior, and the coverage and redundancy of discovery protocols. These findings expose critical differences in network resilience, decentralization, and observability. Beyond characterizing each network, our methodology demonstrates a general framework for measuring decentralized networks at scale. This opens the door for continued monitoring, benchmarking, and more transparent assessments of blockchain infrastructure across diverse ecosystems.

\end{abstract}

\begin{CCSXML}
<ccs2012>
<concept>
<concept_id>10002978.10003029.10011703</concept_id>
<concept_desc>Security and privacy~Usability in security and privacy</concept_desc>
<concept_significance>500</concept_significance>
</concept>
</ccs2012>
\end{CCSXML}

\begin{CCSXML}
<ccs2012>
   <concept>
       <concept_id>10002978.10003014</concept_id>
       <concept_desc>Security and privacy~Network security</concept_desc>
       <concept_significance>300</concept_significance>
       </concept>
   <concept>
       <concept_id>10003033.10003079.10011704</concept_id>
       <concept_desc>Networks~Network measurement</concept_desc>
       <concept_significance>300</concept_significance>
       </concept>
   <concept>
       <concept_id>10003033.10003068</concept_id>
       <concept_desc>Networks~Network algorithms</concept_desc>
       <concept_significance>500</concept_significance>
       </concept>
 </ccs2012>
\end{CCSXML}

\ccsdesc[300]{Security and privacy~Network security}
\ccsdesc[300]{Networks~Network measurement}

\keywords{blockchains, peer-to-peer, network measurements}

\maketitle

\section{Introduction}

    Decentralized systems rely on robust and efficient peer-to-peer (P2P) networking layers to propagate transactions, blocks, and state updates across thousands of participants. In blockchain networks, the P2P layer is foundational not only for consensus and availability but also for performance, censorship resistance, and resilience to churn. Despite its importance, the structure and behavior of blockchain P2P networks remain poorly understood, due in part to the lack of sustained, large-scale measurement across multiple chains.
    
    However, a better understanding is critical. On the one hand, recent regulatory proposals assess \emph{scale} and \emph{control} in crypto networks (e.g., activity thresholds under EU MiCA~\cite{zetzsche2021markets,cointelegraph_defi_eu_2024} and decentralization tests in U.S. proposals~\cite{oranburg2025clarity}). While statutes rarely directly key in on ``node count'', regulators and risk assessors use network size and concentration as evidence of resilience or potential governance capture. 
    On the other hand, due to the permissionless nature of the protocol, it is difficult for protocol designers and maintainers themselves to understand the size and characteristics of the network~\cite{bostoen2024estimating}. For example, a good understanding of the location of nodes plays a crucial role in the design and performance of next-generation blockchain protocols~\cite{alpenglow2025, sousa2015separating}. 
    
    In this paper, we present the first longitudinal, multi-chain measurement study of blockchain P2P networks, covering 36 networks including Bitcoin, Ethereum, Solana, and Dogecoin, among others. From early 2025 onward, we operated 15 dedicated crawlers, incorporated external data from one community crawler, and implemented hourly reachability checks (TCP probes and in-protocol handshakes). To complement this, we performed full IPv4 address space scans (and partial IPv6) for selected protocols using custom payloads and scraped six node explorers to compare crawler coverage. Notably, by analyzing the Ethereum (via discv4 and discv5) and the Bitcoin Cash discovery data, we inferred the existence and metadata of 19 additional networks.

    Our methodology enables a comprehensive view of these P2P layers. We then build on this foundation to analyze structural properties such as churn, geographic and AS-level centralization, peer table characteristics, and inter-network overlap. Together, these analyses uncover substantial diversity in design, scale, and decentralization across blockchain networks. We reveal stark differences in network size (ranging from tens to tens of thousands of nodes), reachability (some networks exhibit <10\% reachable nodes), and peer discovery coverage (how much of the network is revealed by querying a small subset of nodes). We explore IP and AS-level centralization, the use of IPv6, temporal churn behaviors, and quantify peer discovery effectiveness. For example, we find that in some networks, fewer than five nodes need to be contacted to discover 90\% of reachable peers, while others exhibit much flatter coverage.

    We begin with an overview of the peer discovery process of the blockchain systems we study in Section~\ref{sec:background}, and our methodology for our 15 network crawlers (summarized in Table~\ref{tab:summary-protocols}) and general data collection in Sections~\ref{sec:methodology} and ~\ref{sec:datacollection}. We present results from our diverse suite of measurement techniques in Section~\ref{sec:results-size}, culminating in network size estimates for all 36 blockchain networks. We then provide extensive analysis of properties of these networks in Section~\ref{sec:results-analysis}, where we have the data. We conclude in Section~\ref{sec:discussion}. All measurement code will be made publicly available to the community.

    \subsection{Related work}
    \label{sec:relatedwork}

        Network measurements of blockchain systems have primarily focused on the Bitcoin and Ethereum blockchains. Decker et al.~\cite{decker2013information} were the first to measure Bitcoin, focusing on message propagation in the network. Other network propagation measurements of Bitcoin include~\cite{croman2016scaling,pappalardo2018blockchain,neudecker2019short}. Several measurements of Bitcoin have looked at node properties such as size, location, and churn~\cite{imtiaz2019churn,donet2014bitcoin,mariem2020all,park2019nodes}. There is also work on methodologies to estimate the size of unreachable nodes in Bitcoin~\cite{grundmann2022short,wang2017towards} (estimated at two-thirds of the network).
        
        Gencer et al~\cite{gencer2018decentralization} measured peers and propagation in Bitcoin and Ethereum, with a focus on decentralization metrics, while Kiffer et al.~\cite{kiffer2021under} focused on the impact of network churn on information quality in the Ethereum network. Previous measurements of Ethereum have looked at the topological properties of the peer tables~\cite{gao2019topology}, and very recent works have measured the discovery protocols of Ethereum and their clients~\cite{luo2025unveiling,li2025place}.
        
        Several works take advantage of network-level messages to infer peering connections in Bitcoin and Ethereum~\cite{delgado2019txprobe,miller2015discovering,neudecker2016timing,li2021toposhot}. Others have studied peer discovery in Bitcoin and Ethereum, looking at Eclipsing attacks~\cite{heilman2015eclipse,henningsen2019eclipsing,marcus2020low}, AS-level partitioning attacks \cite{saad2019partitioning,apostolaki2021perimeter}. Heimbach et al.~\cite{heimbach2025deanonymizing} ran network-level measurements of Ethereum's Consensus network, looking at general decentralization metrics, and running an attack to extrapolate which nodes participate in the consensus protocol.

        Measurements of other blockchain networks are sparse. 
        Early work (from 2016), which crawled Ethereum, Bitcoin, Peercoin, and Namecoin \cite{anderson2016new},  Cao et al.~\cite{cao2020exploring} crawled Monero, and Daniel et al.~\cite{daniel2019map} Z-cash. There are several works looking at the Lightning Network (a layer-two of Bitcoin) topology, which is publicly advertised \cite{casas2021light,rohrer2019discharged,martinazzi2020evolving}.
        Prior to the age of cryptocurrencies, several works measured the P2P networks of the early 2000s -- such as Gnutella, Kad, BitTorrent, and several P2P-TVs -- looking at similar properties such as size, churn, and traffic~\cite{stutzbach2006understanding,sen2002analyzing,ruckert2014clubbing,bermudez2011passive,ciullo2008understanding}.

\section{Background}
\label{sec:background}

    Unlike traditional client-server architectures, P2P networks rely on a flat topology where nodes act as both clients and servers, directly exchanging information and resources. Efficient peer discovery is required to maintain robustness, scalability, and security in cryptocurrency networks because these networks are generally designed to be open and permissionless, and can undergo heavy churn. Peer discovery protocols determine how nodes learn about and connect to others, directly impacting consensus, data propagation~\cite{kiffer2021under}, and resistance to attacks~\cite{heilman2015eclipse,henningsen2019eclipsing,apostolaki2021perimeter}.
    To facilitate this, many blockchain networks employ variants of structured or unstructured discovery mechanisms, many inspired by distributed hash table (DHT) protocols. Across these mechanisms, a node typically maintains a ``peer table'' or routing table that organizes known nodes according to identifiers such as node IDs, IP addresses, or connectivity metrics. These tables represent a partial view of the network, whose sizes vary significantly by protocol (from tens to over a thousand). 
     
    Although there are many blockchain protocols, the client code for many of them is not entirely unique. 
    Bitcoin's code base has been forked and reused in many existing systems, including Dogecoin, Litecoin, and Bitcoin Cash,\footnote{Bitcoin Cash is not only a code fork, but it used to be part of the Bitcoin blockchain, so it is in fact a blockchain fork.} among others. All these inherit the same peer table and peer discovery mechanism of Bitcoin. There are several protocols, including Ethereum, Polkadot, and Filecoin, among others, whose peer table structure and peer discovery mechanism are based on the Kademlia Distributed Hash Table (DHT) \cite{maymounkov2002kademlia}.
    Other protocols still share some characteristics. We summarize the key characteristics of the protocols studied in this paper below. 

    We omit \textbf{Cardano} below as until recently, their networking stack did not offer peer discovery, and instead relied on static topology files obtained from a centralized source~\cite{cardano2025topology}. Currently, Cardano relies on trusted relays to obtain peer information, with steps towards decentralized discovery in the works~\cite{hryniuk2023dynamicp2p}.

    \subsection{Peer tables}

        Each node keeps some local information of the network, usually in the form of a \textit{peer table} or \textit{routing table}. This is generally not a table of who they are actively connected to, i.e., have an established TCP connection with, but rather a sample of the network they can reference to try and establish a connection to when they need a new peer. 

        \paragraph{Bitcoin-based clients} \textbf{Bitcoin} and its variants (in this work this includes \textbf{Dogecoin}, \textbf{Bitcoin Cash}, and \textbf{Litecoin}) keep track of two peer tables. The \texttt{new} tables stores recently-propagated IPs which can hold up to 66K addresses, while the \texttt{tried} table stores IPs the node has previously established a connection with and holds around 16K addresses. In Bitcoin, there is no peer ID, all nodes are referenced solely by their IP:port information. The peer table does keep some connectivity information (e.g., time since last connection, time in table) which it uses to eject old peers when new ones are added. Periodically (every two minutes) a peer is chosen from the \texttt{new} table, and is promoted to the tried table if the node is able to make a short-lived \textit{feeler} connection to it. When a node needs a new connection, it samples randomly from both tables. More information on peer table maintenance can be found in~\cite{tran2021routing}.

        \paragraph{DHT-based clients}  While various Distributed Hash Table (DHT) designs exist, this study focuses exclusively on Kademlia-based DHTs~\cite{maymounkov2002kademlia}, specifically the discv4~\cite{ethereum_discv4_spec}, discv5~\cite{ethereum_discv5_spec}, and libp2p~\cite{libp2p_go_kad_dht} implementations. In Kademlia, derived from a public key, each peer is assigned a unique node/peer ID within a shared fixed-size XOR keyspace. The protocol defines distance as the bitwise XOR of two node IDs, enabling efficient and symmetric routing. Each node maintains a routing table organized into $k$-buckets, where each bucket covers a range of XOR distances that grow exponentially with respect to a node's ID. This structure ensures that routing queries proceed toward peers that are increasingly closer to a target ID, achieving logarithmic lookup complexity. Despite implementation differences, the analyzed systems share these core Kademlia properties and can be treated within a common analytical framework.
        
        \paragraph{Others} Of the other protocols studied in this work, many implement a simpler peer maintenance mechanism than the ones described above. The common approach is to keep a general cache of discovered nodes (IP:port and public ID) with some protocol-specific additional information for pruning/cleaning the table. In the \textbf{Ripple} protocol, each entry has a hop counter for if the node is a direct peer or peer of a peer, etc. and peer tables have a default size of 256 nodes \cite{xrpl_peerfinder}. In \textbf{Avalanche}, nodes keep some priority information (e.g., if the peer is a validator, successful connection attempts, and their subnets) with default size of between 2000–5000 nodes, depending on release \cite{avalanchego_network}. Both the \textbf{Stellar} and \textbf{Near} protocols nodes keep a default of 1K peers in their discovery cache and similar connectivity metrics (e.g., number of successful attempts, last connection, etc.) \cite{stellar_overlay,stellar_overlay_impl,near_network_architecture}.
            
    \subsection{Peer discovery}

        Periodically nodes ask their peers to whom they have an active connection for some new node information so they can learn about more nodes in the network and update their peer tables. This process also varies by network, including the number of nodes returned, how they choose which information to share, and if the querying client can probe the discovery mechanism with any input.

        \paragraph{Bitcoin-based clients} 
        Here peer discovery is done with a \texttt{getAddress} message which takes no input. The peer returns with up to 1000 nodes sampled from the two peer tables ramdomly (biased by the size of the tables and the freshness of information), this response is cached per IP for 24hrs so that multiple \texttt{getAddress} messages return the same value \cite{bitcoin_addr_response_cache}.

        \paragraph{DHT-based clients} The original Kademlia protocol defines four messages of which the \texttt{FIND\_NODE} message facilitates peer discovery~\cite{maymounkov2002kademlia}. With this request a peer can query any compatible network participant to obtain a list of peers from its routing table that are closest in terms of XOR distance to a specified key. The specific contents of the message varies by implementation. In the discv4 protocol, the query contains the target key of which the requestor wants to get the closest peers known to the remote. The same is true for the libp2p protocol. In the discv5 protocol, the requestor calculates the distance of the remote to the target key and transmits the distance value as part of the \texttt{FIND\_NODE} message. Each of these implementations returns a fixed number of peers per response: 16 in discv4 and discv5, and 20 in libp2p. By carefully crafting these \texttt{FIND\_NODE} messages it is possible to fetch the entire contents of a remote's peer table.

        \paragraph{Others} The other protocols considered in this study have similar discovery mechanisms. The peer discovery protocol of \textbf{Ripple} uses a discovery message with no input. The peer responds with a small sample (typically 10-20 nodes) which is first filled by the node's own peers and then drawn from the cache (node information includes a counter which signals if the peer is actively connected to the node)~\cite{xrpl_peer_protocol}. \textbf{Avalanche} has a discovery mechanism with a mix of push and pull, where you periodically advertise a peer list (default size of 50) made up of first the nodes you are connected to, and the rest of the space with nodes from your discovered cache~\cite{avalanche_peer_list_gossiping}. Both of these protocols prioritize sharing information about their \textit{active} connections. The \textbf{Near} discovery protocol similarly gives information about active connections, where in fact the request contains two fields \texttt{max\_peers} which is how many peers to respond with (up to 50) and \texttt{max\_direct\_peers} which is how many responses should be from active connections, from these they chosen randomly~\cite{near_network_proto}. \textbf{Stellar}, on the other hand, has a peer discovery mechanism which returns a default 50 peers randomly sampled from their peer table~\cite{stellar_overlay_impl}. 
           
    \subsection{Other details}

        \paragraph{Bootstrapping} When a node first joins the network, they may already know about some nodes in the network and first establish a connection to them. Otherwise, a node utilized some \textit{bootstrapping} mechanism to learn about some initial nodes in the network such as some hardcoded nodes in the client or DNS-like servers. In~\cite{loe2019you}, the authors provide a survey of bootstrapping approaches used by several blockchain protocols. 

        \paragraph{Self Announcements} In some networks, nodes periodically gossip an \textit{announcement} of themselves so that other nodes may add them to their peer tables. In Bitcoin-based clients, a node does this via unsolicited \texttt{addr} messages containing their own IP \cite{bitcoin_wiki_satoshi_node_discovery}. In Ethereum's discv4 protocol, the \texttt{FINDNODE} message with the sender's own public ID works also as a self announcement as it initiates the receiver adding the node to their own table if there is space (i.e., the corresponding bucket is empty or has a non-responding entry).

\section{Crawler Methodology}
    \label{sec:methodology}

    \begin{figure}[t]
          \centering
          \includegraphics[width=1\linewidth]{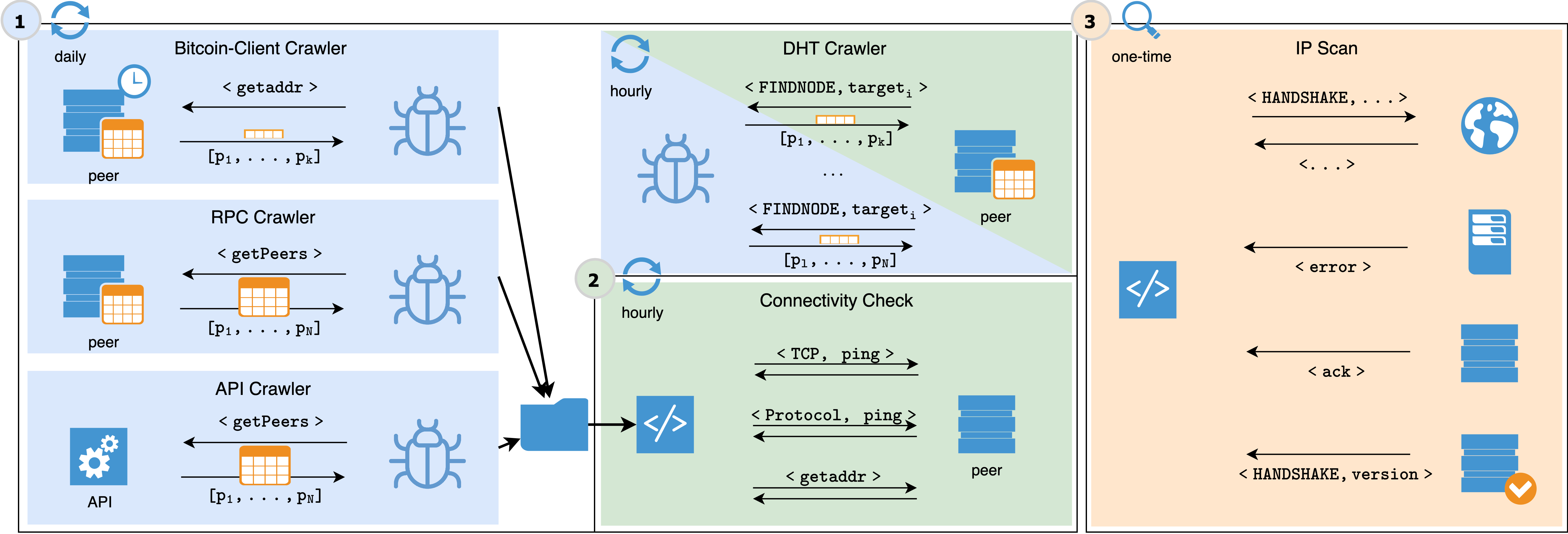}
          \caption{
                Overview of our crawling and connectivity measurement pipeline. 
                (1)~\textit{Crawlers:} For Bitcoin-based, RPC-based, and API-based networks, we query peers for their known peers using messages such as \texttt{getaddr} or \texttt{getPeers}, collecting peer tables daily. For DHT-based protocols, we issue \texttt{FINDNODE} requests hourly to exhaustively enumerate remote routing tables.
                (2)~\textit{Connectivity checks:}  Discovered peers are further probed using a TCP-level connectivity check, as well as a protocol-specific \texttt{PING} message where it exists. Otherwise, the peer discovery mechanism serves as ping (\texttt{crawl PING}).
                (3)~\textit{IP scanning:} For networks lacking peer discovery mechanisms (e.g., Cardano), we perform one-time Internet-wide scans using protocol-specific handshake payloads to identify responsive nodes. 
                Together, these components enable comprehensive and continuous measurement of peer-to-peer blockchain networks across diverse discovery mechanisms.}
        \label{fig:overview}
        \vspace{-5pt}
    \end{figure}
        
    To gather data from a network we primarily utilize the peer discovery mechanism of the network. At a high-level we want to gather the sum of all peer tables, and further distinguish those nodes that are actually online and reachable (we can establish some form of connection to, i.e., are not behind a NAT or firewall). To do this, there are two parts to our data collection: the \textit{crawler} and the \textit{connectivity checker}. Figure~\ref{fig:overview} summarizes our methodology.

    \paragraph{Crawler} We begin with an initial set of nodes we know about,\footnote{For the first crawl, we use hard-coded boostrap nodes, and for subsequent crawls we use a small subset of the nodes we learned about in the previous crawl.} and add them to our queue of nodes to query. A node is then popped from this queue, and a crawl query (typically one or more peer discovery messages) is sent to it. Any new peers learned from the responses are added to the queue if they have not been seen before. This continues until the queue is empty, meaning all nodes have been queried. These requests are executed in parallel, and all responses, along with the peers they originated from, are logged.

    \paragraph{Connectivity Checker} We use several metrics for connectivity checks. One method is to take nodes that responded to our crawl messages. Since a node may be online but not necessarily respond to the crawler for various reasons (e.g., some crawl messages require initiating a peer connection), we use other metrics to establish if a node may be active. One approach is to use in-protocol \texttt{PING} or similar light-weight messages (e.g., a status message) to verify a client is running at the given IP. We also utilize \texttt{TCP PINGs} to test for port-specific activity --- many protocols run on non-standard ports which shouldn't otherwise be open. Since we have various forms of connectivity checks (full crawls, in-protocol \texttt{PING}, \texttt{TCP PING}) to be as inclusive as possible, we generally consider a node "active" if it responds to any of these.\footnote{Because our methodology is entirely active — we initiate all connections — nodes behind NATs or firewalls that do not accept inbound connections are not captured in our active set; thus, our active counts represent a lower bound on true network size. Previous work has estimated the size of “unreachable” nodes in the Bitcoin network as twice the reachable count ~\cite{grundmann2022short,wang2017towards}.} In Table~\ref{tab:daily_network_summary} we show the breakdown of number of nodes that are discovered per crawler and subsequently respond to any of the connection attempts done for that network.

    We now describe the specific messages and approaches used for each network for the crawl and connectivity checker.

    \subsection{Bitcoin-based networks}
        \paragraph{Crawler} We build a crawler based on the Bitcoin Core client~\cite{bitcoincore} which is the reference implementation of Bitcoin based in C++. We use the \texttt{getAddress} message which includes the protocol version and network secret corresponding to each network. Thus to crawl any network running off of a Bitcoin client code fork, all we need is the network secret and initial nodes to bootstrap the crawler. The network secret for Bitcoin is \texttt{0xd9b4bef9}, for Dogecoin it is \texttt{0xc0c0c0c0}, for Bitcoin Cash it is \texttt{0xe8f3e1e3}\footnote{Bitcoin SV and eCash use the same network secret as Bitcoin Cash, and thus effectively run over the same peer discovery network.\textbf{ We therefore refer to the data collected here as BitcoinCash (*) and in Section ~\ref{sec:bitcoincashstar} breakdown the data into the sub-networks based on the version and client information.}}, and for Litecoin it is \texttt{0xdbb6c0fb}. We attempt each connection up to 10 times with a timeout of 20 seconds, and run the connections concurrently. We also send a \texttt{version} message to each discovered node which they respond with protocol-specific details such as the version of the protocoland the client they are running.
        
        \paragraph{Connectivity} Of the nodes discovered that day, we then run hourly connectivity checks which includes an in-protocol \texttt{PING} and a \texttt{TCP PING}. We ping each IPv4 and IPv6 IP address on the port of the address we learned about and on the default port(s) of the protocol (8333 for Bitcoin and Bitcoin Cash (*),  22556 for Dogecoin, and 9333 for Litecoin).
        
    \subsection{DHT-based networks}
        \paragraph{Crawler} We have built a generic Kademlia-DHT crawler based on the Golang discv4~\cite{ethereum_discv4_spec}, discv5~\cite{ethereum_discv5_spec}, and libp2p~\cite{libp2p_go_kad_dht} implementations which was initially employed in~\cite{trautwein2022}. We use the aforementioned \texttt{FIND\_NODE} message with carefully crafted payloads to fetch the contents of a remote peers' individual $k$-buckets. In the discv4 protocol, the query must be constructed such that the target key, when Keccak256-hashed, falls into a target bucket. Similarly, libp2p requires the pre-image to be crafted such that its SHA-256 hash maps to a specific bucket in a remote peers' routing table. In contrast, discv5 allows the requester to directly specify the target distance in the \texttt{FIND\_NODE} request. discv5 and libp2p can span different DHT networks as they have a notion of a protocol ID. Only peers advertising the same protocol ID participate in the same DHT network. discv4 does not have such a concept. One such protocol ID in libp2p is, e.g., \texttt{/celestia/celestia/kad/1.0.0} for the Celestia network and \texttt{0x646973637635} in discv5 (reads \texttt{discv5} in ASCII). Upon a successful connection our crawler performs protocol-specific handshakes to extract additional information about the remote peer. In libp2p it's the \texttt{identify} exchange and in discv5 it's the request for the latest Ethereum Node Record (ENR). This gives us information about, e.g., the software version that the remote peer is running, other supported protocols, or, in the case of discv4 and discv5, which network fork the remote peer follows.
   
        \paragraph{Connectivity} We run a full crawl every two hours for all the studied DHT networks. We tally \texttt{TCP PINGs} if the crawler was able to successfully establish a TCP or QUIC connection to the remote peer. If we further were able to perform the full handshake we label this as a crawl \texttt{PING}.

    \subsection{RPC-based networks}
        Some protocols have built-in Remote Procedure Calls (RPC) where in the node responds with their full current list of active connections. These have the disadvantage that we don't learn about the discovery mechanism of the protocol, but we get exactly the active nodes with much less complexity than setting up a custom discovery messages-based crawler. Both Ripple, Avalanche, and Near expose such functionality.

            \paragraph{Ripple} For daily crawl, we take advantage of a built-in \texttt{crawler} RPC~\cite{xrpl_peer_crawler} in which a node returns all their active connections.\footnote{Except those that are specifically \textit{private} connections, which is a flag a peer can request you set.} We send an \texttt{HTTPS} based request with \url{https://{address}:{port}/crawl} starting with the \url{r.ripple.com} boot service, and then for each IP we learn about we query it on default ports 2459 (IANA assigned) and 51235 (legacy). For connectivity checks, we send a \texttt{health} RPC message and a \texttt{TCP PING}.
            
            \paragraph{Avalanche} Clients have a built-in RPC request \texttt{info.peers}~\cite{avalanche_info_peers} where nodes return all their current active connections. The response includes client version for itself and its peers, the node ID, and some other protocol information (e.g., subnets). The node information includes both an ``IP'' and ``public IP'' which may have different ports, we use the latter. We then run hourly \textit{TCP PINGs} at the given port and default ports 9650 and 9651.  

            \paragraph{Near} Clients have an RPC call \texttt{network\_info} which a node returns all of their active peers (just IP and node ID)~\cite{near_rpc_network}, we run this hourly. Additionally, for the connectivity check, we use a \texttt{status} message and a \texttt{TCP PING} hourly at the given port and default ports 3030 and 24567.
            
    \subsection{Other networks}
            
            \paragraph{Stellar} We adapted the Javascript-based Stellarbeat crawler \cite{stellarbeat_node_crawler}. It uses the discovery messages to enumerate over all discovered peers and ask them for peers, recording one response per node (with some peer information such as if they are a validator).  We run this hourly and record responsive nodes as a \texttt{crawl PING}, plus an hourly \texttt{TCP PING} at the given port and default port 11625.
            
            \paragraph{Solana} Since Solana does not have a traditional peer discovery mechanism, we do not implement our own crawler but instead rely on data from the Sola Beach node explorer \cite{solanabeach}. At the start of our measurement period, their API had a \texttt{/v1/non-validators} which returned the IPs (and some protocol-specific information like version and features) of the RPC nodes of the network. We pulled this data hourly and send a \texttt{TCP PING} to the IPs we learned about at the given port and default ports  8000 and 8001. We note that this API call stopped returning IP-level information in mid May 2025.

            \paragraph{Cardano} Currently, Cardano does not offer a distributed peer discovery mechanism and instead relies on a network of trusted nodes for bootstrapping. In addition, the network consists of different types of nodes (block-producing nodes and relays~\cite{cardano2025docs}), which makes it difficult to measure its size. Therefore, we rely on IP scans, as described in Section~\ref{sec:ipscans}. The scan results are validated through an API maintained by the Cardano foundation~\cite{cardano2025stats}. 

\section{Data Collection}
\label{sec:datacollection}
    \input{figures/overview_table}

    In Table~\ref{tab:summary-protocols} we summarize the networks for which we collect data, sorted by their market caps taken on December 24, 2024. We note that Avalanche was number 9 by market cap at the time. We thus have some level of coverage for 8 out of the top 10 protocols by market cap at the time (Binance was number 5, which we cover in our discv4 breakdown in Section~\ref{sec:ethereumnets} and Figure~\ref{fig:discv4}). Overall, we cover 14 out of the top 21 protocols by market cap (all those above \$5B) and 21 others from the shared nature of several discovery protocols (we list these individually by their respective discovery protocols in Section~\ref{sec:multinetwork}). This makes up the single most extensive coverage of the P2P networks of existing blockchain systems.\footnote{A dashboard with up-to-date results of our active crawls can be found at \url{https://p2pobservatory.networks.imdea.org/}, including references to all our data collection source code. The code for the DHT-based crawlers can be found at \url{https://github.com/dennis-tra/nebula}. Additionally, access to the data will be made available by request for academic use.}

    \subsection{Logistics}
    We run all data collection on AWS in Virginia (us-east-1). The DHT measurements were done every two hours using AWS Fargate serverless compute engines. Depending on the network being crawled the resource configurations differ and range from 1vCPU/2GB in the case of Celestia to 8vCPU/16GB for Ethereum discv4.
    All other crawlers and connectivity checks ran on a 4vCPU/16GB EC2 machine with the Bitcoin-based crawler starting 5 minutes after midnight UTC with each protocol crawled sequentially.
    All other crawls were similarly run sequentially, starting at 01:30 (giving the Bitcoin crawlers plenty of time to finish). The connectivity checkers for these protocols were run hourly starting at 03:00 to 23:00, with the Bitcoin-client checks at the start of the hour, and the others at quarter past. Protocols with hourly crawls were run at this time as well. The node explorer scrapers ran hourly, on their own light 1vCPU/1GB machine. 

    We note that running all measurements from a single geographic location introduces potential biases. Some nodes may apply IP-based rate limiting, filtering, or blocklists targeting cloud provider ranges such as AWS, which could prevent successful handshakes or skew connectivity measurements. Latency between our vantage point and remote peers may also cause occasional timeouts; however, we mitigate this to some extent by considering a node “active” if it responds to any of our hourly checks within a day. Despite this, geographic and provider-based effects remain a limitation and may lead us to undercount certain regions or network segments.
        
    \subsection{Scraping node explorers}
        In Table~\ref{tab:summary-protocols}, we also summarize the networks for which there exists a node explorer (a website that lists information about the nodes running in a network, often run by third parties). More commonly, these websites list general network statistics, such as size and geographic distribution. Some of these explorers explicitly expose the IP addresses of the nodes, most commonly as a snapshot in time (i.e., the current set of nodes in the network with no historical information). Thus, we gather these snapshots over time by scraping the website for the full set of nodes currently active (or calling the API if one exists), and do so on an hourly basis to achieve coverage similar to our crawler collection. 

        For Bitcoin, Bitcoin Cash, Litecoin, and Dogecoin, we use the Blockchair \cite{blockchair} API, which provides a list of IP:port, version, and country-level information. Notably, Blockchair only provides IPv4 data even though all these networks support IPv6. For Bitcoin, we additionally scrape Bitnodes~\cite{bitnodes}, which provides IPv4, IPv6, and Onion addresses for nodes operating over Tor. Bitnodes provides similar information, along with details on how long they have been connected to the node. For the Ethereum execution network, we were scraping Ethernodes~\cite{ethernodes}, which used to provide a list of IP:port, node ID, and client information, as well as the time of last connection. We scraped this information hourly and took nodes whose last connection was from the same date. For Ethereum Classic, we scrape ETCnodes~\cite{etcnodes} hourly, which provides similarly IP:port, node ID, version information, and we filter for nodes whose last seen data was on the same day. Lastly, for Stellar, we use the API call from OBSRVR Radar~\cite{obsrvrRadar} to query snapshot data for historical information (all days of our study) and within a snapshot filter for nodes seen that day.
        
    \subsection{Internet-wide scanning}
        In addition to crawling the cryptocurrency networks, we also perform active measurements to characterize cryptocurrency networks through an additional lens.
        We use ZMap~\cite{durumeric2013zmap} to perform port scans and then leverage ZGrab2~\cite{zgrab}.
        For IPv4, we perform Internet-wide measurements, whereas for IPv6 --- due to its expansive address space --- we rely on the IPv6 Hitlist~\cite{gasser2018clusters} to obtain likely responsive addresses.
        We perform these scans from an AWS machine and follow Internet measurement best practices~\cite{dittrich2012menlo,partridge2016ethical} by using a blocklist, describing our experiments on a website, and limiting our scanning rate to 50kpps.

        We run active measurement scans for three of the Bitcoin-based networks (Bitcoin, Dogecoin, and Bitcoin Cash), and compare the results with ground truth crawling data we collect.
        Additionally, we run two scans of the Cardano network --- a network we do not crawl --- and compare our results with network statistics from the Cardano Foundation. We explain our payload and present the results of this scan in Section~\ref{sec:ipscans}.

\section{Mapping the Landscape: Measuring Peer-to-Peer Blockchain Networks}
    \label{sec:results-size}

    In this section, we describe the diverse set of measurement techniques we deploy to estimate network sizes across 36 blockchain networks. These include active crawlers, decomposing measurements of networks that share a discovery layer, Internet-wide scanning, and comparisons with public node explorers. We 
    synthesize their results into a unified estimate of network sizes. 

    \subsection{Crawler measurements}
    \begin{figure}[t]
        \centering
        \includegraphics[scale=1]{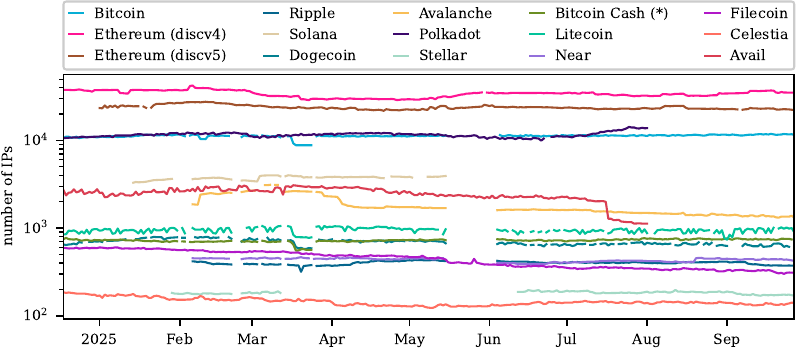 }
        \caption{Daily count of active nodes per blockchain network. Data gaps reflect days with missing observations. The slight drop in March for the Bitcoin-based networks is due to a week in the measurement period where IPv6 was blocked due to a misconfiguration. Occasional gaps in the data reflect crawler connectivity interruptions, including a longer disruption for Avail and Polkadot at the end of the study.}
        \label{fig:active_nodes_per_day}
        \vspace{-10pt}
    \end{figure}
    
    We begin our analysis by presenting an overview of the crawled networks and their temporal evolution. Figure~\ref{fig:active_nodes_per_day} highlights the wide variation in network scale. Ethereum (discv4) and Ethereum (discv5) consistently register over 20,000 active nodes, with Ethereum (discv4) exceeding 30,000 at its peak,\footnote{The large disparity of numbers between the two Ethereum networks is due to their discovery layers encapsulating many other networks, particularly discv4 as we show in Section~\ref{sec:multinetwork}.} while Bitcoin and Polkadot remain in the same order of magnitude. Avalanche, Solana, and Avail sustain a few thousand peers, whereas Litecoin, Stellar, Near, Ripple, Dogecoin, Bitcoin Cash (*), Filecoin, and Celestia operate at only a few hundred. Despite these differences, node counts remain remarkably stable over the eight‑month measurement window.

    Table~\ref{tab:daily_network_summary} reports median daily node metrics across all crawls, where ``active IPs'' denotes peers that replied to our liveness probes and ``discovered IPs'' denotes the total addresses observed in the discovery phase. First, the discovery‑hit ratio (active/discovered) varies dramatically. Polkadot, Filecoin, Celestia, and Avail all exhibit ratios near unity, indicating that almost every discovered peer responds to probes. Thus, nodes in these networks appear to overwhelmingly have active peers in their tables. By contrast, Bitcoin, Dogecoin, Bitcoin Cash, and Litecoin show ratios below 0.05, reflecting large numbers of nonresponsive addresses. Ethereum (discv4) and Ethereum (discv5) fall in between, with ratios of 0.58 and 0.28, respectively.

    Next, we examine hosting versus ISP deployment. Ripple, Solana, Stellar, and Near have hosting‐provider proportions above 0.90, consistent with residential nodes being rare --- especially on resource‑intensive chains like Solana. Bitcoin nodes show the highest ISP share at 0.45, while most other networks lie between 0.10 and 0.40. Further, IPv6 adoption remains limited. Bitcoin‑based networks lead with roughly 15–20\% using IPv6, whereas most chains rely almost entirely on IPv4; notable exceptions are Polkadot (0.20) and Celestia (0.16).

    Finally, liveness‑check success varies markedly by probe type. TCP‑level pings achieve median success rates above 0.80 for most networks; notable exceptions are several DHT-based networks: Polkadot (0.19), Ethereum (discv4) (0.66), Ethereum (discv5) (0.62), and Avail (0.18). Each blockchain is probed at a higher level via either protocol or crawl pings. Bitcoin client–based crawlers employ protocol pings, yielding median response rates above 0.80.
    Nebula DHT–based crawlers use crawl pings, which succeed at median rates above 0.90 for Ethereum (discv4), Ethereum (discv5), Polkadot, Filecoin, Celestia, and Avail. In general, Bitcoin-based networks were more likely to respond to TCP pings while DHT-based networks were more likely to respond to crawl pings. 
    
    \begin{table}[t]
    \resizebox{\textwidth}{!}{
    \begin{tabular}{@{}l|rrr|rr|rr|rrr@{}}
    \toprule
     & active IPs & discovered IPs & A/D ratio & prop. ISP & prop. hosting & prop. IPv4 & prop. IPv6 & prop. TCP ping & prop. protocol ping & prop. crawl ping \\
    \midrule

    Bitcoin & 11309 & 257019 & 0.04 & 0.46 & 0.53 & 0.85 & 0.15 & 0.96 & 0.87 & 0.00 \\
    Ethereum (discv4) & 33318 & 57806 & 0.57 & 0.17 & 0.82 & 1.00 & 0.00 & 0.70 & 0.00 & 0.91 \\
    Ethereum (discv5) & 21685 & 80797 & 0.27 & 0.23 & 0.76 & 0.94 & 0.06 & 0.62 & 0.00 & 0.90 \\
    Ripple & 404 & 891 & 0.45 & 0.08 & 0.90 & 1.00 & 0.00 & 1.00 & 0.94 & 0.00 \\
    Solana & 3693 & 4589 & 0.81 & 0.04 & 0.96 & 1.00 & 0.00 & 1.00 & 0.00 & 0.00 \\
    Dogecoin & 695 & 14945 & 0.05 & 0.38 & 0.61 & 0.82 & 0.18 & 0.99 & 0.92 & 0.00 \\
    Avalanche & 1581 & 5290 & 0.28 & 0.22 & 0.77 & 1.00 & 0.00 & 0.91 & 0.00 & 0.00 \\
    Polkadot & 9151 & 9386 & 0.97 & 0.43 & 0.55 & 0.80 & 0.20 & 0.19 & 0.00 & 0.93 \\
    Stellar & 182 & 1118 & 0.16 & 0.05 & 0.94 & 1.00 & 0.00 & 0.22 & 0.00 & 1.00 \\
    Bitcoin Cash (*) & 725 & 27328 & 0.03 & 0.24 & 0.75 & 0.81 & 0.19 & 0.96 & 0.82 & 0.00 \\
    Litecoin & 952 & 28063 & 0.03 & 0.24 & 0.75 & 0.80 & 0.20 & 0.95 & 0.90 & 0.00 \\
    Near & 435 & 777 & 0.54 & 0.02 & 0.98 & 1.00 & 0.00 & 0.85 & 0.28 & 0.15 \\
    Filecoin & 411 & 417 & 0.99 & 0.51 & 0.41 & 0.92 & 0.08 & 0.86 & 0.00 & 0.97 \\
    Celestia & 138 & 144 & 0.96 & 0.20 & 0.78 & 0.84 & 0.16 & 0.86 & 0.00 & 1.00 \\
    Avail & 2224 & 2244 & 0.99 & 0.38 & 0.61 & 0.89 & 0.11 & 0.18 & 0.00 & 0.99 \\
    \bottomrule
    \end{tabular}}\vspace{2pt}
    \caption{Median daily node metrics for each blockchain: active, discovered node counts, the discovery‐hit ratio (active/discovered), and median proportions of nodes by network attributes (ISP vs.\ hosting), IP version (IPv4 vs.\ IPv6), and liveness checks via TCP, protocol, and crawl pings.}
    \vspace{-20pt}
    \label{tab:daily_network_summary}
    \end{table}

        \subsection{Multi-network discovery protocols}
            \label{sec:multinetwork}
            
            While many blockchain networks operate isolated peer discovery mechanisms, others share a common discovery layer across multiple logical protocols or chains. In such cases, a single discovery network may simultaneously serve multiple sub-protocols, forks, or application-layer overlays. This multiplexing complicates network measurement and attribution, as the discovery layer does not always explicitly reveal which protocol a peer belongs to. In this section, we examine three such discovery networks, Bitcoin Cash (*), and Ethereum’s discv4 and discv5 discovery mechanisms, and decompose them into their distinct protocols where possible. Beyond these shared discovery networks, we also quantify IP-level overlap across all 14 measured discovery mechanisms to capture cross-network participation in Appendix~\ref{app:overlap}.
            
            \subsubsection{Bitcoin Cash network}
                \label{sec:bitcoincashstar}
                
\begin{figure}[t]
                  \centering
                  \begin{minipage}[t]{0.35\textwidth}
                    \vspace{9pt}
                    \centering
                    {\scriptsize
                    \begin{tabular}{lrr}
                      \hline
                      \textbf{\makecell{Protocol}} & \textbf{\makecell{Market Cap \\ (12/24)}} & \textbf{\makecell{Network Size\\(IPv4/IPv6)}}  \\
                      \hline
                      BitcoinCash   & \$11.9B & 289/89 \\
                      BitcoinSV & \$1.6B &  28/4  \\
                      eCash    & \$0.73B &  133/36 \\
                      none     & X & 155/24 \\
                      other &    X &  1/0 \\
                      \hline
                    \end{tabular}
                    }
                  \end{minipage}
                  \hfill
                  \begin{minipage}[t]{0.64\textwidth}
                    \vspace{0pt}
                    \centering
                    \includegraphics[width=\linewidth]{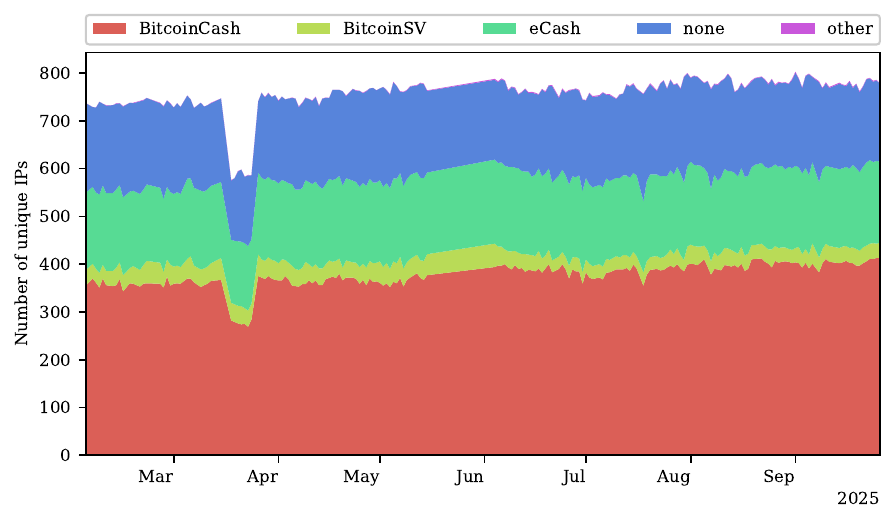}
                  \end{minipage}
                  \caption{Networks sharing the same peer discovery as Bitcoin Cash. On the right, we see the active peer count. Recall that in March there was a period with no IPv6 scanning, which accounts for the less than 10\%  drop.}
                  \Description{}
                  \label{fig:bitcoincash}
                \end{figure}

                In the Bitcoin ecosystem, the network-specific \texttt{MAGIC} string is meant to differentiate networks (including testnets) as an early check if two nodes follow the same blockchain. While investigating the top Bitcoin \textit{forks},\footnote{Here we mean a \textit{blockchain fork} where a network used to be part of the canonical Bitcoin blockchain until protocol changes led them to start building different blocks ~\cite{elman2021new}.} we found that \textbf{Bitcoin Cash} -- a Bitcoin fork since November 2018 -- and \textbf{Bitcoin SV} -- actually a fork of Bitcoin Cash since July 2019 -- shared the same \texttt{MAGIC} string. This means that at the discovery layer, they act as one network, not differentiating until further down in the peering handshake. Thus, they store each other's IPs in their peer tables. To differentiate them, we look at the \texttt{version} message, specifically which \textit{client} a node runs. Through this analysis, we also found a third protocol \textbf{eCash} -- another Bitcoin Cash fork since November 2020 \cite{pontem_ecash2024} -- with its own client. Figure~\ref{fig:bitcoincash} shows the breakdown across the three main networks and the IPs unresponsive to the \texttt{version} message (i.e., the "none" category). For each IP, on a given day, we check \texttt{version} responses up to a week before and after to increase our labeling. Given its market cap, Bitcoin SV makes up a small fraction of our labeling. 
            
        \subsubsection{Ethereum networks}
            \label{sec:ethereumnets}

                \FloatBarrier
                \begin{figure}[t]
                  \centering
                  \begin{minipage}[t]{0.34\textwidth}
                    \vspace{9pt}
                    \centering
                    {\scriptsize
                    \resizebox{\textwidth}{!}{
                    \begin{tabular}{lrr}
                      \hline
                      \textbf{\makecell{Protocol}} & \textbf{\makecell{Market Cap \\ (12/24)}} & \textbf{\makecell{Network Size}}  \\
                      \hline
                      Ethereum   & \$462.8B &  9,203 \\
                      Binance   & \$103.3B & 1,299 \\
                      Gnosis & \$696M & 1,774  \\
                      \textbf{Polygon} & \$1.4B & 1,129 \\
                      ETH Classic  & \$5.3B & 716 \\
                      Syscoin   & \$91.4M &  578 \\
                      Energi   & \$4.83M &  475 \\
                      \textbf{Scroll}   & \$227M &  249 \\
                      \textbf{Linea}   & \$391M &  233 \\
                      Berachain   & \$257M &  347 \\
                      Pulsechain   & \$843M &  209 \\
                      Story   & \$1.29B &  156 \\
                      AirDao   & \$26.0M &  111 \\
                      Vitruveo   & \$4.60M & 191  \\
                      Etho   & \$0.93M &  61 \\
                      Vana   & \$514M & 73  \\
                      \textbf{Metis}   & \$283M &  72 \\
                      Edgeware   & 0 &  36 \\
                      Fuse   & \$6.65M &  21 \\
                      Lumoz   & \$17.4M &  6 \\
                      other   & X & 3,935 \\
                      misc client   & X &  2,540 \\
                      none   & X &  5,719 \\
                      \hline
                    \end{tabular}
                    }}
                  \end{minipage}
                  \hfill
                  \begin{minipage}[t]{0.65\textwidth}
                    \vspace{0pt}
                    \centering
                    \includegraphics[width=\linewidth]{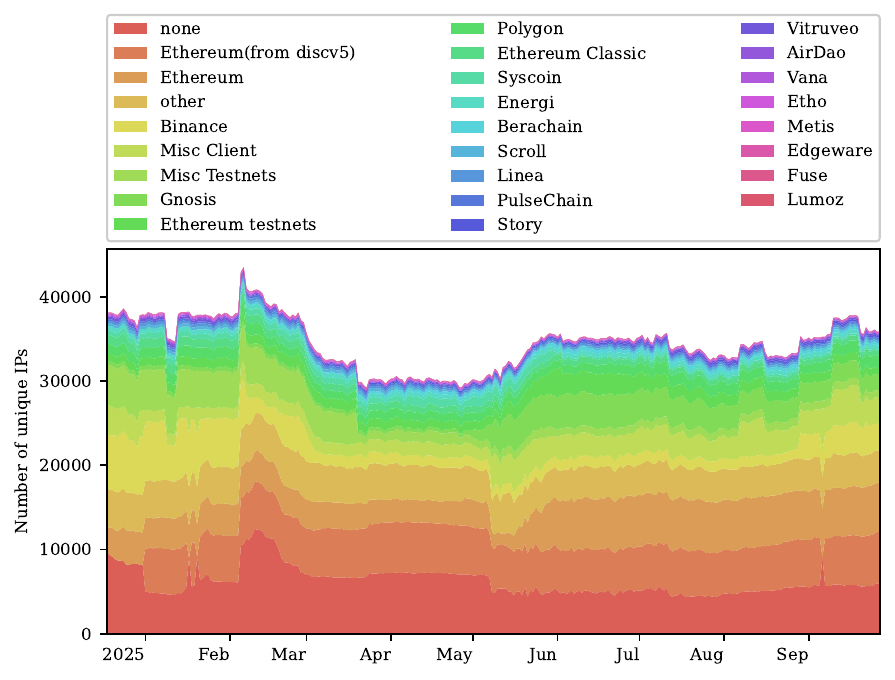}
                  \end{minipage}
                  \caption{We break down the active nodes we learn about in the Ethereum Execution network (those running over discv4), their market cap on December 24, 2024, and their median size in our measurement period. Ethereum Layer-2 networks are in bold (with Linea showing TVL since it doesn't have a native token). Berachain and Story did not exist in December 2024, so we take the market cap at the end of our study period, July 15, 2025. Edgeware ceased to exist in May 2025. }
                  \Description{}
                \label{fig:discv4}
                \end{figure}

                \begin{figure}[t]
                  \centering
                  \begin{minipage}[t]{0.35\textwidth}
                    \vspace{9pt}
                    \centering
                    {\scriptsize
                    \resizebox{\textwidth}{!}{
                    \begin{tabular}{lrr}
                      \hline
                      \textbf{\makecell{Protocol}} & \textbf{\makecell{Market Cap \\ (12/24)}} & \textbf{\makecell{Network\\ Size}}  \\
                      \hline
                      Ethereum   & \$462.8B & 9,913 \\
                      Gnosis & \$696M & 514  \\
                      Holeskey Testnet & X & 721  \\
                      Hoodi Testnet & X & 655  \\
                      Sepolia Testnet & X & 618  \\
                      none & X &  10,377 \\
                      other & X & 1,198  \\
                      \hline
                    \end{tabular}
                    }}
                  \end{minipage}
                  \hfill
                  \begin{minipage}[t]{0.64\textwidth}
                    \vspace{0pt}
                    \centering
                    \includegraphics[width=\linewidth]{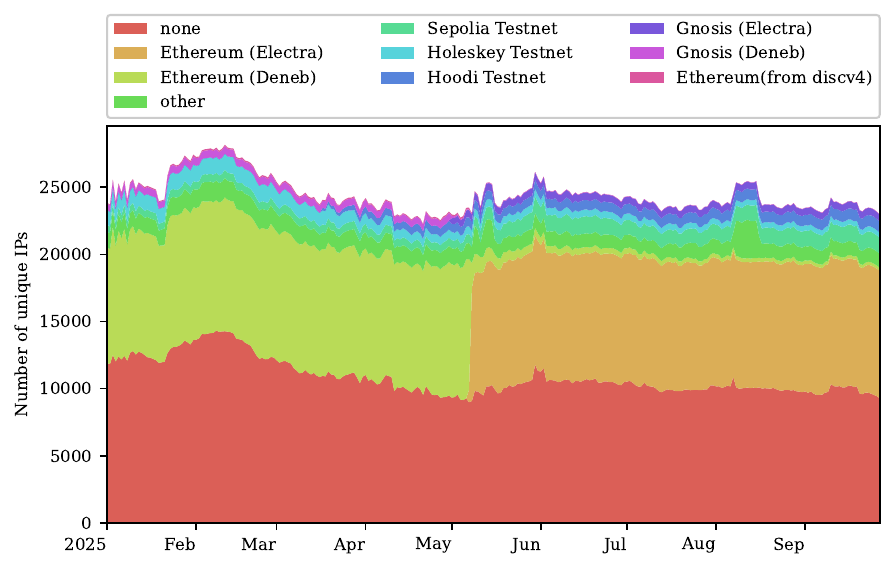}
                  \end{minipage}
                  \caption{We break down the active nodes we learn about in the Ethereum discv5 discovery network.}
                  \Description{}
                  \label{fig:discv5}
                \end{figure}

            Ethereum uses two generations of node discovery protocols, discv4 and discv5, which serve as the foundation for many Ethereum-based networks to find and connect to peers. Unlike Bitcoin-style forks, where blockchains split at the consensus level, Ethereum's discovery layer acts more like a shared substrate: multiple networks (even those not related to the Ethereum ecosystem) may use the same discovery graph to bootstrap connectivity, even if they do not interoperate at higher layers. 

            Recall that in discv4, nodes are identified by a public key (called the \textit{enode}) which does not give any protocol-level information. For this, we initiate a protocol handshake with all nodes we learn about. This handshake provides three fields which we can use to identify a network: \textit{fork id} (a protocol-specific hex, specific to the unique version (or fork) of the protocol), \textit{chain id} (a number to signify the network\footnote{Some are unique to the protocol but others like \textbf{1} is used by the Ethereum mainnet and several other networks. A breakdown of chain ids can be found in \cite{chainid_network}.}) and \textit{client} (mostly unique for protocols running non-Ethereum clients such as the bor client for the Polygon network \cite{polygon_bor}). We label networks first by \textit{fork id}, then by \textit{chain id} if it is unique, and lastly by client (again by uniqueness, then a ``Misc Client'' field for all Ethereum-based clients with no fork id or chain id given). Figure~\ref{fig:discv4} summarizes our labeling with the top protocols and their market caps. As with Bitcoin Cash, we take all unlabeled IPs (from the ``none'' category --- meaning no information is given) and look at a window of one week to see if it can inherit a label from a different day. We also take discv5 Ethereum-labeled IPs and cross-reference them with the ``none'' IPs; these are labeled ``Ethereum(from discv5)''. This decreases the none category by more than half. This implies a difficulty for us to execute the full protocol handshake with half of the Ethereum network peers on the execution network (discv4), while still receiving some response (e.g., a \texttt{too many peers} error) which allows us to verify they are active.  

            For discv5, nodes have ENRs, which encode several protocol-level information with, most importantly for us, the \textit{fork digest}, which encodes both the genesis block data\footnote{The first block of a blockchain which defines the whole chain} and the current protocol version. We can use this exclusively to differentiate the Ethereum mainnet (which underwent a protocol upgrade during our measurement study from the Deneb fork to Electra), its testnets, and, interestingly, the Gnosis protocol (also present in discv4). We have a small ``other'' category with fork digests associated with unknown protocols (possible private networks), and still a large ``none'' for IPs, which gave no ENR information. Again, we take any labelling within a window of one week, but given the encoded information in the ENR, this does not noticeably decrease the number of unlabeled IPs, and cross-referencing the discv4 labeling for Ethereum gives us only a handful of additional labels.

    \subsection{Comparison with node explorers}

\begin{figure}[t]
        \centering
        \includegraphics[scale=1]{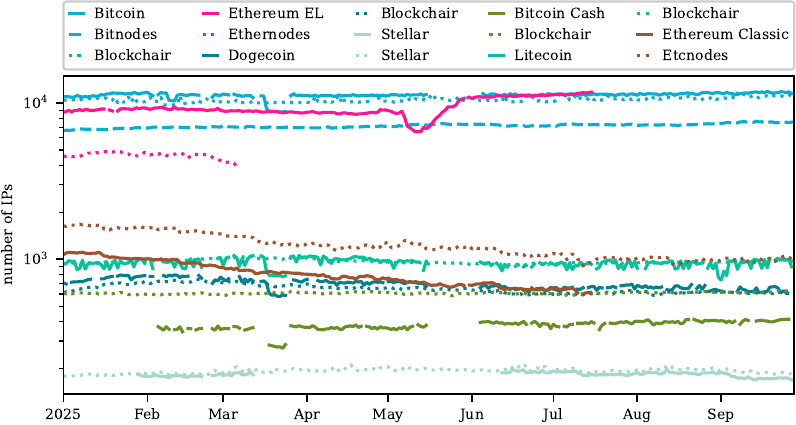}
        \caption{Daily unique‐IP node counts for major blockchains, comparing our crawler data (solid lines) against explorer counts (dotted and dashed lines). Each network is shown in a consistent color. Note that Ethernodes stopped providing IP data in March 2025.}
        \Description{}
        \label{fig:active_nodes_explorer_counts}
    \end{figure}

    Finally, we compare our network size measurements to those of node explorers where possible. Figure~\ref{fig:active_nodes_explorer_counts} contrasts our crawler‑derived active‑IP counts (solid lines) with public explorer data (dotted and dashed lines). For Bitcoin, Dogecoin, and Litecoin, Blockchair’s figures track our measurements closely but remain slightly lower (we attribute this to the missing 10\% of IPv6). Bitnodes reports a markedly smaller Bitcoin network than Blockchair and our measurements. In the case of Bitcoin Cash, Blockchair overestimates active peers by about twofold, likely because it does not differentiate between the multiple blockchains using the same discovery protocol (cf. Section~\ref{sec:bitcoincashstar}).  

    Moving to two blockchains using the Ethereum (disv4) protocol, Ethernodes significantly underestimates node counts versus our crawler (likely because of the missing discv5 information cf. Section~\ref{sec:ethereumnets}), whereas Etcnodes' estimate for Ethereum Classic slightly exceeds ours. Stellar stands out for its near‑perfect alignment between explorer and crawler counts, likely because they also utilize the Stellar Beat crawler implementation \cite{stellarbeat_node_crawler}. Most of these explorers don't provide methodology, so we can only speculate about the discrepancies, but they underscore how differing measurement methods affect reported network sizes.

    \subsection{Internet-wide scanning}
    \label{sec:ipscans}
    Many blockchains use their own client software, so we often had to write new code to handle their specific peer discovery and handshake behavior. To reduce effort and as a potential general solution to gathering network measurements across ecosystems, we explored Internet-wide scans. Here, we probe all IP addresses in the IPv4 space and all addresses present on the IPv6 hitlist~\cite{gasser2018clusters}, initiating a TCP connection at a specific port (e.g., the default port of the protocol) and by using a network-specific payload (e.g., the first exchange of the peering handshake). While the payload does require some technical understanding of the protocol, it is significantly simpler than setting up an entire crawler. 

    To test the efficacy of Internet-wide scanning on capturing actual network participants, we first scan three of the Bitcoin-client based networks (Bitcoin, Dogecoin, and Bitcoin Cash (*)) that we have ground truth data for from our crawls. We additionally scan the Cardano network, the 8th largest by market cap as of December 2024, for which we do not have a working crawler and present the results below. 
    
    \subsubsection{Bitcoin clients IP scan}
    \label{app:bitcoin-scan}
    For the Bitcoin clients, we use the \texttt{VERSION} message with the network-specific \texttt{MAGIC} string. We verify that only clients with the same \texttt{MAGIC} string respond to our request; otherwise, the connection attempt times out. 
    We consider primarily the default port for scanning. Table~\ref{tab:top_ports} shows the port breakdown of the IPv4 active clients in these three networks during this measurement period. We confirm that the default ports are the most widely used.
    For each network, we scan the associated port with the associated payload. For Dogecoin, we also scan one additional port (19919), and for Bitcoin Cash (*) we scan two additional ports (8344 and 8433), to get more coverage since the default ports have slightly lower use.

    We ran the scans consecutively from July 9 to 14, 2025. We take the crawl data for these networks from the day(s) of the scan of the protocol, plus a two-day window (e.g., the Bitcoin scan ran on July 8, so we take crawl data from July 6 to 10) to compare. Table~\ref{tab:bitcoin_scan} summarizes our Bitcoin-based results with the number of successful connections we get, those that respond with some payload, and what was in the payload. For most protocols, we see a large fraction of HTTP 400 error messages, which hints at contacting an HTTP server running on a non-standard port. We check if the payload corresponds to a correct Bitcoin version message and specifically the correct version of the network (Version\_x in the table). Of these, we take the intersection with the discovered and active nodes from our crawler data. 

    We see that the IP addresses we find in scans consistently show up in our active data set, which is expected, as they are reachable. From Table~\ref{tab:top_ports}, we would expect to learn about 10K IPs in the Bitcoin scan, and we do get about 90\% of this number. There is less overlap with our crawl data, which could be due to the random sampling nature of Bitcoin's discovery protocol, or the 5\% daily churn we observed in the data (cf. Section~\ref{sec:churn}). Dogecoin sees a similar percentage of coverage, and the version message is able to distinguish well the nodes actually in the Dogecoin network (the majority of nodes that respond with a version that is not Dogecoin's are nodes with Bitcoin's version number). Bitcoin Cash, on the other hand, uses the same version number as Bitcoin, so its scan responses are much more polluted with Bitcoin data. It is interesting that because these scan payloads used the network-specific \texttt{MAGIC} string, most of the Bitcoin nodes do not respond to the Bitcoin Cash (*) payload. A final note is that less than 5\% of the version-specific responses were from IPv6 addresses (216 for Bitcoin, 12 for Dogecoin, and 14 for Bitcoin Cash (*)). 
    \textbf{The above results suggest Internet-wide scans do give a good estimate of the network size when a protocol has a unique payload and response.}

    \begin{table*}[t]
        \resizebox{\textwidth}{!}{
        \begin{tabular}{l|r|rr|rr|rr}
        \toprule
        \textbf{Network} & \textbf{Ports} & \textbf{Disc. Count} & \textbf{Disc. \%} &
        \textbf{TCP Count} & \textbf{TCP \%} &\textbf{Protocol Count} & \textbf{Protocol \%}\\
        \midrule
        Bitcoin & 8333 & 245.9K & 95.5 & 10.4K & 96.2 & 9.8K & 96.6 \\
        Dogecoin & 22556, 19919 & 11.5K, 522 & 92.7, 4.2 & 499, 43 & 79.0, 6.8 & 462, 42 & 79.1, 7.2 \\
        Bitcoin Cash (*) & 8333, 8334, 8433 & 18.4K, 1.2K, 281 & 67.6, 4.6, 1.0 & 492, 18, 9 & 73.1, 2.7, 2.1 & 382, 18, 13 & 72.1, 3.4, 2.5 \\
        \bottomrule
        \end{tabular}}\vspace{2pt}
        \caption{Top ports for the Bitcoin-based networks by total discovered, those that responded to the TCP ping, and those that responded to the in-protocol ping. Data is taken from the days of the IPv4 scan of each network with a 2-day buffer on either side.}
        \label{tab:top_ports}
        \vspace{-20pt}
    \end{table*}

    \begin{table*}[t]
        \resizebox{\textwidth}{!}{
        \begin{tabular}{l|rrrrr|rrrr}
        \toprule
        \textbf{Network} & \textbf{Success} & \textbf{Response} & \textbf{400 Error} &
        \textbf{Version\_all} & \textbf{Version\_x} & \textbf{Disc.} &\textbf{TCP} & \textbf{Proto}\\
        \midrule
        Bitcoin & 428,751 & 207,453 & 176,569 & 9,427 & 9,093 & 7,486 & 7,432 & 7,431 \\
        Dogecoin & 337,364 & 26,346 & 3,653 & 2,321 & 392 & 282 & 279 & 279\\
        Bitcoin Cash (*) & 607,013 & 206,693 & 182,549 & 2,584 & 1,290 & 238  & 237 & 235 \\
        \bottomrule
        \end{tabular}}\vspace{2pt}
        \caption{Here we show the results of the IPv4 scan measurements. For each network, we show how many successful connections we got and how many of those returned a value in the response. We decode the response and see mostly HTTP 400 error messages of the successful connections and those that can be decoded into the \texttt{VERSION} response, of those, we take the subset that have the network-specific version number. Of this last group, we look at the overlap with our crawl data with those discovered, and those that responded to the TCP and in-protocol Pings. }
        \label{tab:bitcoin_scan}
        \vspace{-20pt}
    \end{table*}

    \subsubsection{Cardano IP scan}
    Since Cardano relies on a network of trusted nodes for bootstrapping~\cite{hryniuk2023dynamicp2p} and consists of multiple types of nodes~\cite{cardano2025docs}, we measure the size of its network solely through IP scans. Cardano's networking layer consists of multiple mini-protocols. The handshake protocol is used to negotiate the protocol version and the protocol parameters when connecting to a node, and consists of a single request and a single reply encoded in CBOR~\cite{cardano2025network}. 
    
    Based on existing client implementations, we craft a valid handshake request that requests establishing a connection using all available protocol versions.
    We scan the five most commonly used ports, according to the ports advertised in the only public registry we could find (ports 3000, 3001, 3002, 6000, and 6001)~\cite{koios2025relays}. Together, these ports make up 75\% of all advertised ports by relays. Each port is scanned sequentially and scans last up to 6 days.
    
    Table~\ref{tab:cardano} shows how many responses we collect. As these ports are shared with multiple other protocols and applications, the vast majority of responses are not valid handshake responses. The valid responses further consist of \texttt{msgRefuse} (code=2) or \texttt{msgProposeVersions} (code=0)~\cite{coutts2025ouroboros}. We consider only handshake successes (\texttt{msgAcceptVersion}, code=1). The vast majority of successful handshake responses accept the latest protocol version (version 14). We validate these results using the Cardano Foundation API~\cite{cardano2025stats}, which posts aggregate statistics on network size. We compare our results with the average network size over the days of the scans. These numbers are slightly lower than our measurements, which can be explained because our scans cannot distinguish testnets or private networks, and the scan collects data over multiple days, which could aggregate some churn.

    \begin{table*}[t]
        \centering
        \resizebox{.95\textwidth}{!}{
        \begin{tabular}{l|cccccc|r}
        \toprule
        \textbf{Scan Date} & \textbf{Success} & \textbf{Response} & \textbf{400 Error} & \textbf{Valid} & 
        \textbf{Handshake} & \textbf{Latest Version} & \textbf{API} \\
        \midrule
        May  & 2,355,577 & 1,618,246 & 1,112,032 & 3,029 & 1,838 & 1,726 & 1,311 \\
        July & 2,402,764 & 1,661,047 & 1,118,110 & 2,925 & 1,784 & 1,689 & 1,285 \\
        \bottomrule
        \end{tabular}
        }
        \vspace{2pt}
        \caption{IP scan results for Cardano. The first column shows the number of unique IP addresses to which we successfully connected on either port 3001, 3001, 3002, 6000, or 60001. Further columns show the subset of IP addresses from which we received a response, those that were HTTP 400 error, a valid response (correctly formatted CBOR), a successful handshake (code = 1), and that support the newest protocol version (version 14). The last column shows the average number of nodes returned by the API over the time of the IP scans.}
        \vspace{-20pt}
        \label{tab:cardano}
    \end{table*}

    \subsection{Network size summary}\label{sec:fulloverview}

\begin{figure}[htbp]
        \centering
        \includegraphics[scale=1]{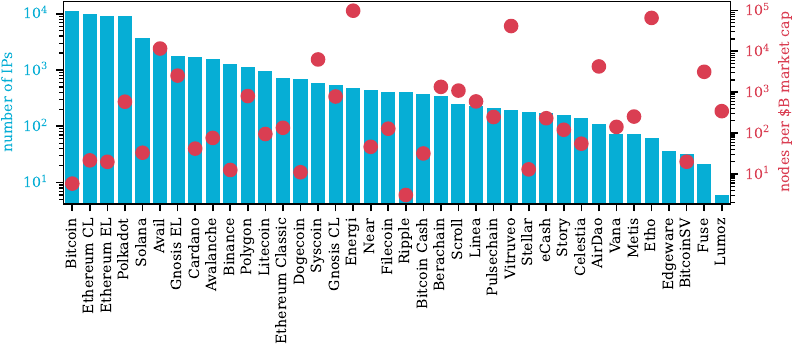}
        \caption{Median network size (number of unique IPs) across blockchain networks. Bars represent each network’s median node‑count, highlighting the wide variations between networks.}
        \Description{}
        \label{fig:median_network_size}
        \vspace{-10pt}
    \end{figure}

    Given our crawls, scans, and breakdown of discovery protocols, we can finally summarize our findings. We display the median network sizes for all crawled blockchains together with their economic scale in Figure~\ref{fig:median_network_size}. \textbf{The analysis underscores the extreme heterogeneity of blockchain network footprints: median peer counts span more than three orders of magnitude, from over 10,000 in the largest networks down to fewer than 10 in the smallest.} We also relate network size to market capitalization (right y-axis), expressing node density as active peers per US\$ billion. While market capitalization is an imperfect baseline as it omits measures such as TVL or on-chain activity, it provides a coarse economic reference. The comparison shows no linear relationship between capitalization and network size but reveals pronounced differences across ecosystems: several small-cap networks maintain relatively dense infrastructures, whereas high-value systems operate with far fewer nodes per unit of capitalization.

\section{Beyond Size: Structural Properties and Decentralization}
    \label{sec:results-analysis}

With network size baselines in place, we now turn to a broader analysis of structural and operational properties that influence decentralization and resilience. We explore temporal dynamics (churn, uptime), geographic and network-layer concentration, peer table characteristics, and inter-network relationships (cf. Appendix~\ref{app:overlap} for full overlap matrices across all networks). We focus primarily on the daily data gathered through the 15 discovery network crawlers in this section. Although the metrics vary by design and maturity of each protocol, together they provide insight into the centralization risks and operational diversity present across blockchain P2P ecosystems.

    \subsection{Temporal dynamics}

    \subsubsection{Daily churn}
    \label{sec:churn}
    
\begin{figure}[t]
        \centering
        \includegraphics[scale=1]{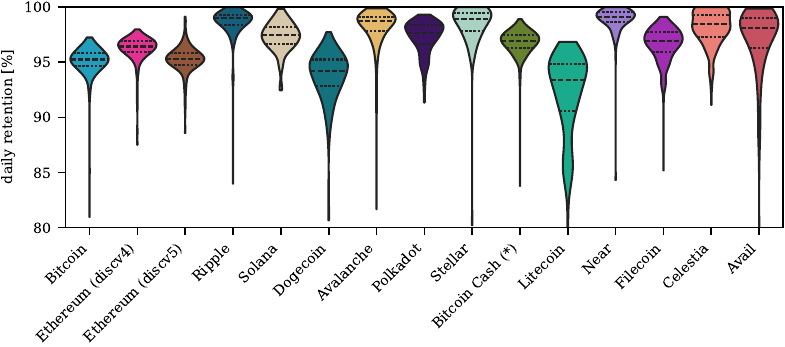}
        \caption{ Distribution of daily IP-retention rates for each of the 15 blockchain networks. Retention on day $t$ is defined as the percentage of IPs active on day $t-1$ that remain active on day $t$. Violin shapes show the full density of daily retention values, with inner quartiles and medians indicated.}
        \Description{}
        \label{fig:retention_violin}
        \vspace{-8pt}
    \end{figure}

    We begin by examining churn via daily node‐retention rates, defined as the percentage of nodes active on day $t-1$ that remain active on day $t$. Figure~\ref{fig:retention_violin} presents the full distribution of these rates for all 15 blockchains, with medians and inner quartiles indicated.

    All networks sustain high retention: every chain has median rates above 90\%. The lowest medians appear in the Bitcoin‑based networks, Dogecoin and Litecoin, which fall below 95\%. In contrast, smaller networks such as Ripple, Avalanche, Stellar, Near, and Celestia achieve medians above 98\%. Notably, the upper tails of several distributions (e.g., Bitcoin and Litecoin) cut off well below 100\%, revealing a baseline level of daily churn that may result from some nodes rotating their IPs. Weekly retention rates, where churn compounds over multiple days, are shown in Figure~\ref{fig:weekly_retention_violin} in Appendix~\ref{sec:app_churn}. We also examine churn from the opposite perspective by considering daily new‑IP rates in Figure~\ref{fig:new_nodes_violin} in Appendix~\ref{sec:app_churn}. Several networks show a nonzero lower bound in their new‑IP distributions. Bitcoin and Litecoin never fall below roughly 2\%, suggesting a steady influx of fresh IPs from address rotation or dynamic hosting. In contrast, networks like Ripple, Avalanche, Near, Celestia, and Avail have median new‑IP rates below 2\%, indicating highly stable peer sets. 
    
\begin{figure}[t]
        \centering
        \includegraphics[scale=1]{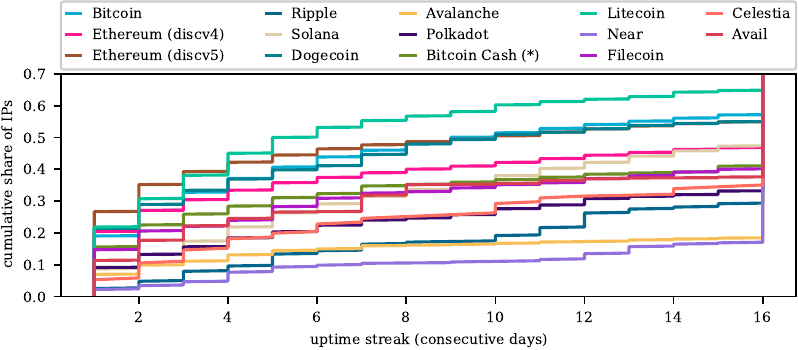}
        \caption{Cumulative distribution of IP uptime streaks across 14 blockchain networks, April 5–20 2025. For each network, the curve shows the cumulative share of IPs (y‑axis) whose observed consecutive‑day uptime streak does not exceed the value on the x‑axis. Note that Stellar is not included in this figure, as we do not have data for the specified window.}
        \Description{}
        \label{fig:uptime_streaks_cdf_april}
    \end{figure}

    Finally, we examine multi‑day persistence by plotting the cumulative distribution of node uptime streaks for the two‑week window from April 5–20, 2025 (Figure~\ref{fig:uptime_streaks_cdf_april}). Each curve shows the fraction of nodes whose observed consecutive‑day uptime does not exceed $x$ days. Smaller chains such as Near and Avalanche display the highest persistence, with the majority of peers active for all 16 days. By contrast, larger or Bitcoin‑based networks --- Bitcoin, Litecoin, Dogecoin, and Ethereum (discv5) --- have fewer than 50\% of nodes sustaining uninterrupted operation over the full period. These results demonstrate that even modest daily churn (see Figure~\ref{fig:retention_violin}) compounds over time, yielding substantial turnover across multi‑day intervals.

    \subsection{Geographic and AS distribution}

    \subsubsection{Geographic distribution}

    \begin{figure}[t]
        \centering
        \includegraphics[scale=1]{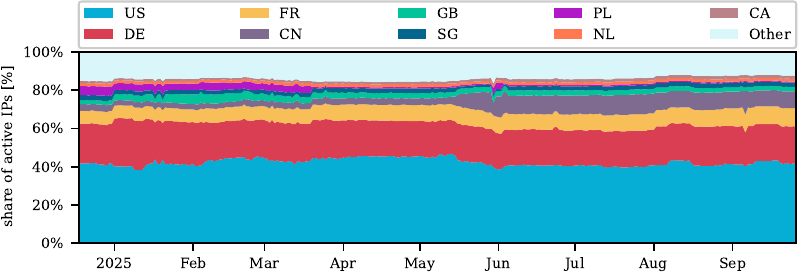}
        \caption{Daily percentage share of active blockchain IPs across all networks by country. The nine countries with the highest node counts, i.e., United States (US), Germany (DE), France (FR), Netherlands (NL), United Kingdom (GB), Switzerland (CH), Poland (PL), China (CN), and Canada (CA), are shown individually, with all other countries aggregated into ``Other''.}
        \Description{}
        \label{fig:all_country}
    \end{figure}

    \begin{table}[t]
        \centering
        \resizebox{0.75\textwidth}{!}{  
        \begin{tabular}{@{}lrrrrrr@{}}
        \toprule
         & Africa & Asia & Europe & North America & Oceania & South America \\
        \midrule
        Bitcoin & 0.3 & 15.3 & 36.7 & 45.3 & 1.4 & 1.0 \\
        Ethereum (discv4) & 0.2 & 7.9 & 40.0 & 50.7 & 0.8 & 0.5 \\
        Ethereum (discv5) & 0.2 & 4.8 & 49.4 & 43.8 & 1.0 & 0.8 \\
        Ripple & 0.2 & 16.6 & 44.3 & 38.6 & 0.2 & 0.0 \\
        Solana & 2.1 & 0.6 & 29.8 & 65.7 & 0.0 & 1.7 \\
        Dogecoin & 0.3 & 8.9 & 46.7 & 42.3 & 1.6 & 0.1 \\
        Cardano & 0.2 & 7.2 & 46.4 & 45.2 & 0.7 & 0.2 \\
        Avalanche & 0.0 & 2.0 & 33.1 & 64.6 & 0.3 & 0.1 \\
        Polkadot & 0.4 & 25.5 & 33.1 & 40.1 & 0.2 & 0.7 \\
        Stellar & 0.0 & 4.9 & 45.9 & 48.6 & 0.5 & 0.0 \\
        Bitcoin Cash (*) & 0.1 & 8.2 & 54.3 & 35.9 & 1.4 & 0.1 \\
        Litecoin & 0.2 & 8.3 & 60.2 & 30.5 & 0.6 & 0.2 \\
        Near & 0.0 & 0.7 & 60.0 & 39.3 & 0.0 & 0.0 \\
        Filecoin & 0.0 & 59.2 & 19.3 & 19.3 & 2.0 & 0.2 \\
        Celestia & 0.7 & 3.5 & 53.1 & 37.1 & 0.7 & 4.9 \\
        Avail & 1.7 & 23.6 & 35.5 & 33.9 & 3.9 & 1.4 \\
        \midrule 
        Overall & 0.3 & 10.6 & 41.0 & 46.5 & 0.9 & 0.7 \\
        \bottomrule
        \end{tabular}}\vspace{2pt}
        \caption{Percentage distribution of active IPs by continent for each blockchain network. Values represent the fraction of daily observed IPs located in Africa, Asia, Europe, North America, Oceania, and South America, respectively. The final ``Overall'' row aggregates these percentages across the 15 crawled networks as well as Cardano.}
        \vspace{-20pt}
        \label{tab:continent_dist}
    \end{table}

    We next examine the geographic distribution of active IPs across all blockchains with IP mappings given by IPInfo~\cite{ipinfo}. Figure~\ref{fig:all_country} shows the daily share of IPs in the nine countries with the highest node counts (United States, Germany, France, Netherlands, United Kingdom, Switzerland, Poland, China, and Canada), with all other countries grouped as ``Other''. The combined share of the United States and Germany remains above 50\% throughout the measurement period, and the top nine countries account for over 80\%  of all active IPs, indicating pronounced geographic centralization. Although these figures aggregate across all networks, Appendix~\ref{app:country} provides the country breakdown for each individual chain. In most cases, the United States and Germany host the largest fractions of peers. Notable exceptions include Polkadot, where China holds the second largest share, and Filecoin, in which China contributes the largest share of IPs.

    Further detail is provided in Table~\ref{tab:continent_dist}, which breaks down each network’s median share of active IPs by continent alongside the overall averages. Overall, Europe and North America dominate, accounting for 41.2\% and 46.5\% of IPs respectively, with Asia contributing another 10.4\%. Three networks -- Polkadot, Filecoin, and Avail -- stand out with over 20\% of peers in Asia, and Filecoin approaches 60\% Asian presence. Solana and Avail also exceed the global average of 0.3\% for Africa, each with more than 1.5\% of nodes on the continent. Aside from Solana (29.8\% Europe) and Filecoin (19.3\% Europe and 19.3\% in North America), every network places at least 30\% of its IPs each in both Europe and North America, with Solana’s node population heavily skewed toward North America (65.7\%). Additionally, Avail shows notable concentrations in Oceania (3.9\%) and Celestia in South America (3.1\%). We show the continent breakdown for blockchains in the multi-discovery networks individually in Appendix~\ref{app:continent}.

    \subsubsection{AS-level distribution}
    We next quantify AS-level concentration with the  Herfindahl–Hirschman Index (HHI). The HHI is defined as
    \[
    \mathrm{HHI}_{\mathrm{AS}} \;=\; \sum_{j=1}^{M} s_j^2,
    \]
    where \(M\) is the number of distinct autonomous systems (ASes) observed and
    \[
    s_j = \frac{\text{number of active nodes in AS }j}{\text{total active nodes}}
    \]
    is the share of active nodes in AS \(j\). For each day \(t\), we compute the daily HHI\(_{\mathrm{AS}}\) by summing the squared shares \(s_j^2\). An HHI\(_{\mathrm{AS}}\) near \(1/M\) indicates low concentration (nodes evenly distributed across ASes), whereas values approaching 1 indicate high concentration in a few ASes.

\begin{figure}[htbp]
        \centering
        \includegraphics[scale=1]{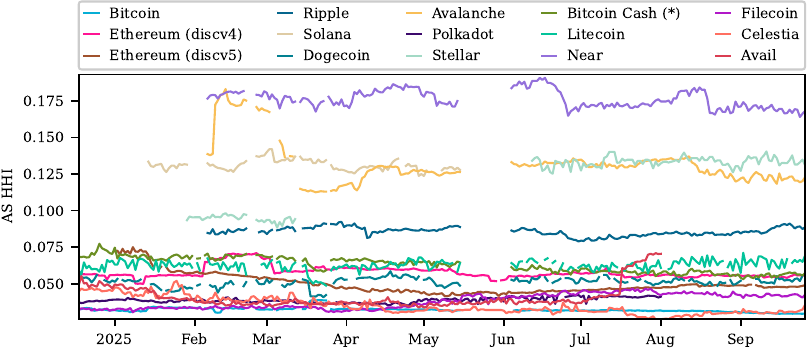}
        \caption{Daily Herfindahl–Hirschman Index (HHI) of Autonomous System for active nodes across 15 blockchain networks. Higher HHI values reflect greater concentration of nodes within fewer ASNs, signalling increased centralization of network hosting.}
        \Description{}
        \label{fig:asn_hhi_all}
    \end{figure}

    Figure~\ref{fig:asn_hhi_all} plots the daily HHI of AS‑level concentration for each of the 15 networks. Several chains exhibit markedly higher concentration, including Near, Avalanche, Ripple, Solana, and Stellar. These networks either have relatively small networks (Near, Ripple, Stellar) or rely heavily on data‑center deployments to meet high resource requirements (Solana), resulting in most peers residing in a handful of ASes. Avalanche stands out in the plot, having experienced a large fluctuation of its HHI, matching a general large fluctuation of network participants during the first couple of months of our study.  
    
    By contrast, Filecoin, Bitcoin, Celestia, Avail, and Polkadot consistently rank among the least concentrated. Bitcoin and Polkadot have large networks, while Avail and Celestia already showed low geographic centralization, and Filecoin’s limited use of large hosting providers further reduces its AS HHI.   

    \subsection{Peer table analysis}

    Until now, we have largely focused our analysis on the active network. We end our analysis by diving deeper into the peer tables to gain an understanding of how much of the network nodes in the network know about and what proportion of peers they tell us about is active. Recall that we do not collect peer tables with our Solana measurements, so it is omitted from this section.
    Table~\ref{tab:peer_stats_median} summarizes four key peer‑table metrics for each blockchain: the median table size, the median count of peers in each table that respond to liveness probes, the proportion of table entries that are active peers, and the fraction of the overall active network covered by the median table.

    \begin{table}[t]
        \resizebox{0.8\textwidth}{!}{
        \begin{tabular}{@{}lrrrr@{}}
        \toprule
         & peer table size & active peers in table & prop. active & prop. covered \\
        \midrule
 
        Bitcoin & 863 & 138 & 0.165 & 0.012 \\
        Ethereum (discv4) & 159 & 153 & 0.986 & 0.004 \\
        Ethereum (discv5) & 194 & 180 & 0.926 & 0.007 \\
        Ripple & 19 & 15 & 0.762 & 0.039 \\
        Dogecoin & 729 & 339 & 0.463 & 0.467 \\
        Avalanche & 1285 & 829 & 0.669 & 0.521 \\
        Polkadot & 108 & 99 & 0.920 & 0.008 \\
        Stellar & 50 & 27 & 0.551 & 0.150 \\
        Bitcoin Cash (*) & 737 & 26 & 0.034 & 0.036 \\
        Litecoin & 723 & 28 & 0.039 & 0.029 \\
        Near & 34 & 19 & 0.543 & 0.045 \\
        Filecoin & 95 & 91 & 0.952 & 0.191 \\
        Celestia & 59 & 53 & 0.889 & 0.373 \\
        Avail & 98 & 90 & 0.923 & 0.036 \\
        \bottomrule
        \end{tabular}}
        \caption{Summary of median peer‑table metrics for each blockchain: median table size; median count of active peers per table; fraction of peers in each table that are active; and the share of the active network covered by the median table.}
        \vspace{-20pt}
        \label{tab:peer_stats_median}
    \end{table}

    First, table sizes vary by more than an order of magnitude. Avalanche returns the largest tables (median size = 1349 peers), which is distinctly surprising since the data we collect for Avalanche is their \textit{active} connections. The Bitcoin-based protocols by default return up to 1k peers, which are generally below this -- Bitcoin (864) and Bitcoin Cash(*) (744), and Dogecoin (718). At the other extreme, Ripple returns only 19 peers on average, with Near (34), which makes sense since the data we collect is their active connections. Stellar (50) and Filecoin (97) also return fewer. 

    Next, we see stark differences in peer responsiveness. Ethereum (discv4) and Ethereum (discv5) achieve over 93\% active peers in their tables, as do Polkadot (92\%), Filecoin (95\%), Celestia (89\%), and Avail (92\%), reflecting highly well‑maintained peer lists in these DHT-based networks. In contrast, the Bitcoin-based networks have few active peers in their tables. To be precise, Bitcoin, Bitcoin Cash (*), and Litecoin tables have fewer than 5\% active entries, possibly due to a previously measured 2/3 of the Bitcoin network being behind NATs \cite{grundmann2022short}. Though Ripple, Avalanche, and Stellar nodes return their active connections, the low response rate of those connections to our crawls (less than 30\% in any network) makes it challenging to extrapolate meaningful topological properties of these networks. For the other networks, it is difficult (by design) to infer active connections from peer table data, as not all active connections are stored in the peer table, and peer tables often have many more nodes than connections. Further analysis is left for future work. 

    Finally, table coverage of the full network ranges from under 1\% for both Ethereum networks (0.4\% and 0.7\%) up to over 50\% for Avalanche, whose median table alone sees more than half of all active peers. Dogecoin (39\%) and Celestia (38\%) also achieve substantial coverage, whereas most other networks cover below 10\% of their active nodes with a single crawl response.

\begin{figure}[t]
        \centering
        \includegraphics[scale=1]{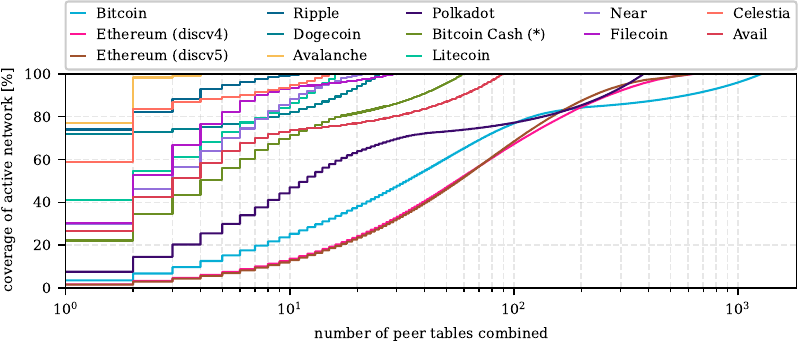}
        \caption{Greedy coverage of active IPs across discovery networks on April 10, 2025. Each curve shows the cumulative percentage of active IPs covered as additional peer tables are greedily unioned, ordered to maximize new node discovery at each step. The x‐axis (log scale) is the number of peer tables combined, and the y‐axis is the percentage of active IPs reached. Note that Stellar is not included in this figure, as we do not have data for the day.}
        \Description{}
        \vspace{-10pt}
        \label{fig:greedy_coverage_all_chains}
    \end{figure}

    We further assess peer‑table coverage using a greedy union algorithm on April 10, 2025 (Figure~\ref{fig:greedy_coverage_all_chains}). Each curve traces the cumulative share of active IPs reached as additional peer tables are added in order of maximal marginal gain. Avalanche’s first table already discovers over 75\% of active IPs, and just four tables are required to cover the entire network. A single table from Ripple, Dogecoin or Celestia also suffices to reach at least 50\% of active IPs. In contrast, larger networks demand far more tables to get to all active IPs: Bitcoin needs 1,278 peer lists to cover all active nodes, Ethereum (discv4) requires 640 lists, Ethereum (discv5) 623 lists, and Polkadot 380 lists. Further, a single peer list for each of these networks does not suffice to reach 10\% of the active network. 

\section{Concluding Discussion}
\label{sec:discussion}

    Our findings reveal a surprising gap between the economic weight of many blockchain ecosystems and the size and transparency of the P2P networks that sustain them. Some networks supporting millions—or even billions—of dollars in value are maintained by only a few dozen publicly reachable nodes. This fragility has real security implications: small or centralized networks may be easier to partition, censor, or disrupt. Furthermore, many users and developers remain unaware of the actual state of the networks they rely on. For several systems, especially those lacking public node explorers, there is no simple way to even approximate network size or health. This lack of visibility undermines trust and makes it difficult to assess the decentralization or resilience that these systems often claim to embody.

    Our measurement effort also underscores just how difficult it is to observe blockchain P2P infrastructure at scale. Each network tends to implement its own custom discovery logic, often with sparse or outdated documentation and code written in disparate languages. Even for this study, which covers a representative set of major cryptocurrencies by market cap, we required more than a dozen custom-built crawlers and extensive protocol analysis. In some cases, default port scanning offered a scalable alternative, but only when protocol signatures were consistent and predictable. Importantly, because our methodology is entirely active, we only capture nodes that are publicly reachable; peers behind NATs or firewalls are not reflected in our “active” counts, which therefore represent a lower bound on true network size. Passive measurements could complement this view, but would require running full nodes across multiple networks, which is substantially more resource-intensive. Developers could further improve transparency by including basic protocol metadata earlier in the handshake, adopting network-unique identifiers (e.g., MAGIC bytes), documenting peer table maintenance strategies, and building public-facing measurement tools. Notably, many node explorers today are maintained by independent third parties rather than protocol developers themselves -- a gap that speaks to a broader neglect of observability at the network layer, the critical foundation supporting all higher layers of the trillion‑dollar cryptocurrency ecosystem.

\bibliographystyle{ACM-Reference-Format}
\bibliography{references}
\appendix

\newpage
\section{Ethics} 

This study follows the Menlo Report’s~\cite{dittrich2012menlo} four principles for ethical ICT research: \emph{Respect for Persons}, \emph{Beneficence}, \emph{Justice}, and \emph{Respect for Law and Public Interest}. We evaluate risks ex ante, minimize potential harms during data collection and analysis, and practice transparency and accountability.
Our datasets deal exclusively with publicly available data, crawls take up minimal connections (connect, ask for info, disconnect), and each node is connected to at a maximum once every one or two hours.
Our measurements target publicly reachable blockchain nodes rather than individuals. No demographics or behavioral data about people were collected. Internet Protocol (IP) addresses can be considered personal data in some jurisdictions; we therefore treat them as sensitive technical identifiers.

\section{Network Churn}
    \label{sec:app_churn}

    \begin{figure}[htbp]
        \centering
        \includegraphics[scale=1]{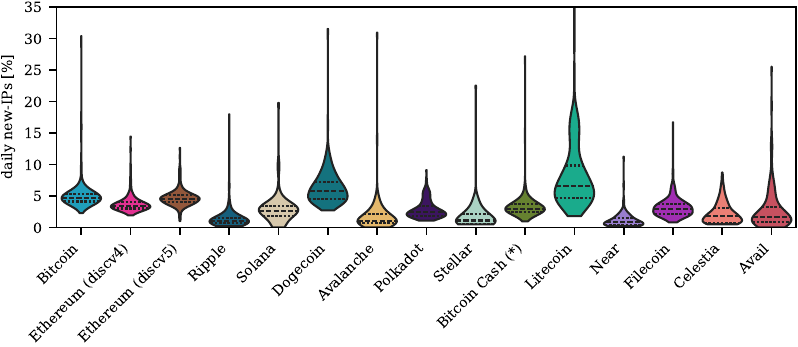}\vspace{-5pt}
        \caption{ Distribution of daily new‑IP rates for each of the 15 blockchain networks. The new‑IP rate on day $t$ is defined as the fraction of IPs active on day $t$ that were not active on day $t-1$, expressed as a percentage of the day $t-1$ active‑IP count. Violin shapes show the full density of daily retention values, with inner quartiles and medians indicated.}\vspace{-5pt}
        \Description{}
        \label{fig:new_nodes_violin}
    \end{figure}

Figure~\ref{fig:new_nodes_violin} presents the distribution of the fraction of IPs active on day $t$ that were not active on day $t-1$, expressed as a percentage of the day $t-1$ active‑IP count. Several networks exhibit a nonzero lower bound in their new‑IP distributions. For example, Bitcoin and Litecoin never drop below roughly 2\% -- suggesting a persistent influx of fresh IPs, perhaps due to address rotation or dynamic hosting. In contrast, smaller or more specialized networks such as Ripple, Avalanche, Near, Celestia, and Avail display low new‑IP rates (medians below 2\%), reflecting highly stable networks.

    \begin{figure}[htbp]\vspace{-3pt}
        \centering
        \includegraphics[scale=1]{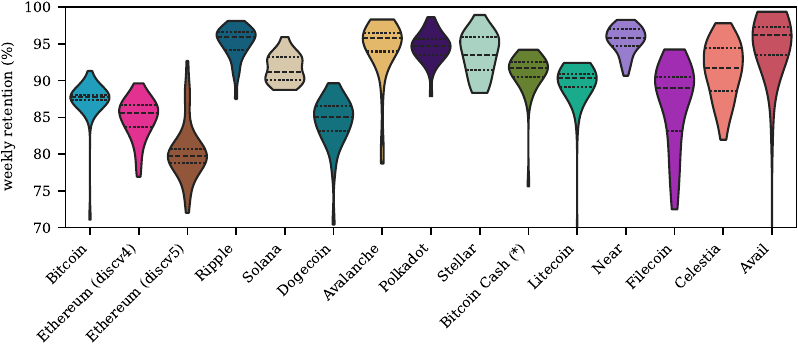}\vspace{-5pt}
        
        \caption{Distribution of daily new‑node percentages across the 15 blockchain networks. Each violin shows the full density of the fraction of nodes joining on a given day (new nodes today as a percentage of the previous day’s total), with inner quartiles and medians marked.}\label{fig:weekly_retention_violin}
        \Description{}
    \end{figure}

We further examine the weekly IP retention rates in Figure~\ref{fig:weekly_retention_violin}. The distributions show that no network retains 100\% of nodes over the span of a week. While daily retention is generally extremely high (cf. Figure~\ref{fig:retention_violin}), the effect of network churn compounds over several days. Ethereum (discv5) has the lowest median retention at around 80\%, whereas Avail shows the highest with a median above 95\%. Most networks cluster above 90\%, with larger networks such as Bitcoin, Ethereum (discv4 and discv5), and Dogecoin falling slightly below that threshold.

\section{Country Split}\label{app:country}

\begin{figure}[H]\vspace{-5pt}
    \centering
    \begin{subfigure}{0.48\linewidth}
    \includegraphics[scale=1]{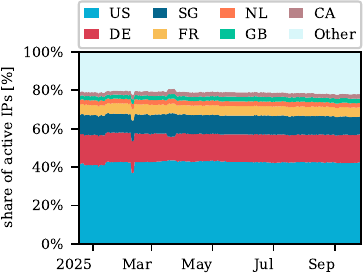}
    \caption{Bitcoin}\label{fig:bitcoin_country}
    \end{subfigure}\hfill
    \begin{subfigure}{0.48\linewidth}
    \includegraphics[scale=1]{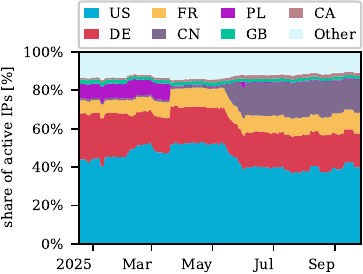}
    \caption{Ethereum (discv4)}\label{fig:ethel_country}
    \end{subfigure}\vspace{-0pt}

    \begin{subfigure}{0.48\linewidth}
    \includegraphics[scale=1]{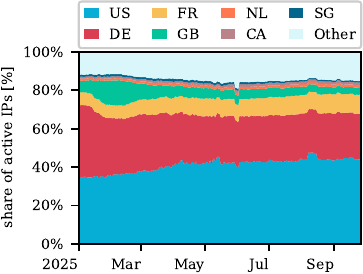}
    \caption{Ethereum (discv5)}\label{fig:ethcl_country}
    \end{subfigure}\hfill
    \begin{subfigure}{0.48\linewidth}
    \includegraphics[scale=1]{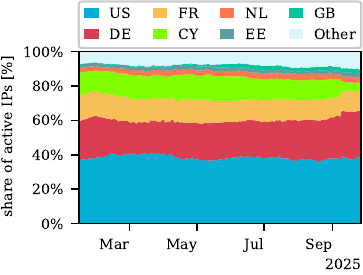}
    \caption{Ripple}\label{fig:xrp_country}
    \end{subfigure}\vspace{-10pt}
    
    \caption{Daily percentage share of active blockchain nodes by country for each blockchain network. The seven countries with the highest node counts are shown individually, with all other countries aggregated into ``Other''.}
    \Description{}
    \label{fig:country_by_network}
\end{figure}

\FloatBarrier

\begin{figure}[H]\ContinuedFloat
    \begin{subfigure}{0.48\linewidth}
    \includegraphics[scale=1]{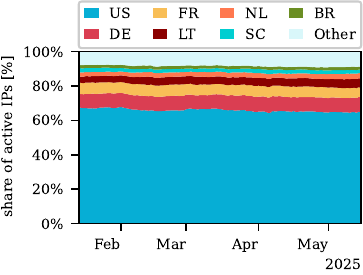}
    \caption{Solana}\label{fig:solana_country}
    \end{subfigure}\hfill
    \begin{subfigure}{0.48\linewidth}
    \includegraphics[scale=1]{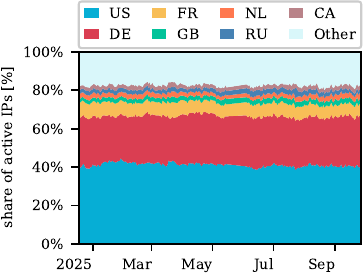}
    \caption{Dogecoin}\label{fig:dogecoin_country}
    \end{subfigure}

    \begin{subfigure}{0.48\linewidth}
    \includegraphics[scale=1]{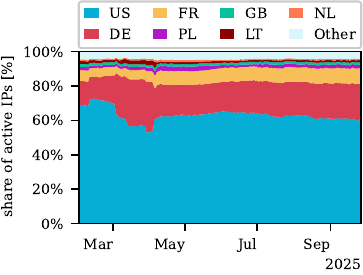}
    \caption{Avalanche}\label{fig:avalanche_country}
    \end{subfigure}\hfill
    \begin{subfigure}{0.48\linewidth}
    \includegraphics[scale=1]{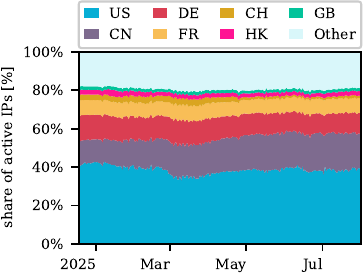}
    \caption{Polkadot}\label{fig:polkadot_country}
    \end{subfigure}

    \begin{subfigure}{0.48\linewidth}
    \includegraphics[scale=1]{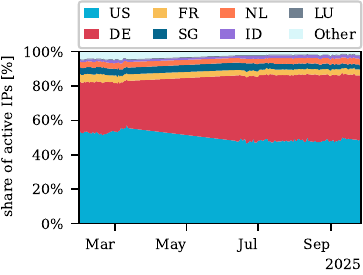}
    \caption{Stellar}\label{fig:stellar_country}
    \end{subfigure}\hfill
    \begin{subfigure}{0.48\linewidth}
    \includegraphics[scale=1]{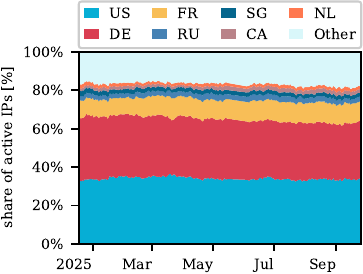}
    \caption{Bitcoin Cash (*)}\label{fig:bitcoincashsv_country}
    \end{subfigure}

    \caption{Daily percentage share of active blockchain nodes by country for each blockchain network. The seven countries with the highest node counts are shown individually, with all other countries aggregated into ``Other''.}
    \Description{}
    \label{fig:country_by_network_2}
\end{figure}

We provide the country‑level breakdown for each individual network in Figure~\ref{fig:country_by_network_3}. In most cases, the United States and Germany host the largest fractions of peers, as expected. Notable exceptions include Polkadot, where China holds the second‑largest share, Filecoin, in which China contributes the largest share of IPs, and Near, where Germany surpasses the United States as the primary host of peers.

\begin{figure}[H]\ContinuedFloat
    \begin{subfigure}{0.48\linewidth}
    \includegraphics[scale=1]{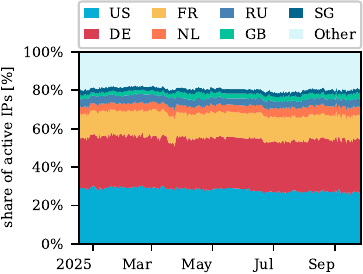}
    \caption{Litecoin}\label{fig:litecoin_country}
    \end{subfigure}\hfill
    \begin{subfigure}{0.48\linewidth}
    \includegraphics[scale=1]{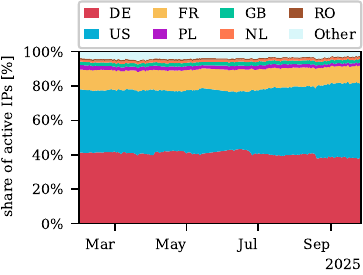}
    \caption{Near}\label{fig:near_country}
    \end{subfigure}
    
    \begin{subfigure}{0.48\linewidth}
    \includegraphics[scale=1]{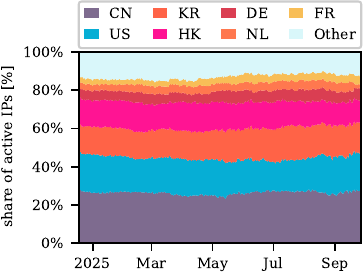}
    \caption{Filecoin}\label{fig:filecoin_country}
    \end{subfigure}\hfill
    \begin{subfigure}{0.48\linewidth}
    \includegraphics[scale=1]{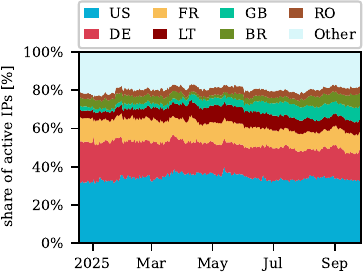}
    \caption{Celestia}\label{fig:celestia_country}
    \end{subfigure}

    \begin{subfigure}{0.48\linewidth}
    \includegraphics[scale=1]{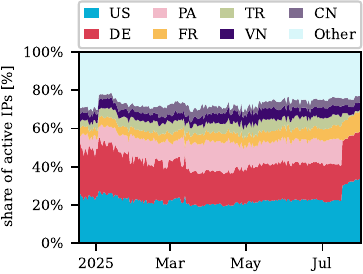}
    \caption{Avail}\label{fig:avail_mainnet_country}\vspace{-10pt}
    \end{subfigure}
    
    \caption{Daily percentage share of active blockchain nodes by country for each blockchain network. The seven countries with the highest node counts are shown individually, with the rest aggregated into ``Other''.}
    \Description{}
    \label{fig:country_by_network_3}
\end{figure}

\section{Continent Split}\label{app:continent}
Table~\ref{tab:continent_dist_all} reports the geographic distribution of active nodes per network, with blockchains in the multi-network discovery protocol broken down. 
The breakdown highlights differences across multi-network deployments: for instance, in the Ethereum (discv4) discovery protocol, Etho shows a strong concentration in Europe, whereas Edgeware has only a small presence there. Similarly, eCash exhibits higher activity in Asia and lower representation in North America compared to the two other blockchains in the Bitcoin Cash (*) discovery protocol.

\FloatBarrier
\begin{table*}[htbp]
\centering
\resizebox{\textwidth}{!}{
\begin{tabular}{l l rrrrrr}
\toprule
\textbf{Group} & \textbf{Network} & Africa & Asia & Europe & North America & Oceania & South America \\
\midrule

\multirowcell{1}{\rotatebox[origin=c]{90}{}} 
 & Bitcoin & 0.3 & 15.3 & 36.7 & 45.3 & 1.4 & 1.0 \\
 \midrule

\multirowcell{16}{\rotatebox[origin=c]{90}{\shortstack[c]{Ethereum\\(discv4)}}}

 & Ethereum (from discv5) & 0.1 & 5.4 & 49.0 & 43.3 & 1.7 & 0.5 \\
 & Ethereum& 1.1 & 18.0 & 33.6 & 45.0 & 1.6 & 0.7 \\
 & Binance & 0.1 & 7.0 & 49.0 & 43.5 & 0.2 & 0.2 \\
 & Gnosis & 0.2 & 6.1 & 38.1 & 53.6 & 1.6 & 0.5 \\
 & Polygon & 0.0 & 2.4 & 43.1 & 54.4 & 0.1 & 0.0 \\
 & Ethereum Classic & 0.3 & 9.9 & 55.2 & 33.2 & 1.1 & 0.3 \\
 & Syscoin & 0.0 & 2.7 & 25.9 & 71.4 & 0.0 & 0.0 \\
 & Energi & 0.2 & 15.5 & 54.7 & 28.9 & 0.4 & 0.2 \\
 & Berachain & 0.3 & 0.3 & 44.3 & 54.6 & 0.3 & 0.3 \\
 & Scroll & 0.0 & 3.1 & 37.1 & 59.4 & 0.4 & 0.0 \\
 & Linea & 0.0 & 1.7 & 58.1 & 40.2 & 0.0 & 0.0 \\
 & PulseChain & 0.0 & 0.5 & 68.7 & 28.4 & 2.4 & 0.0 \\
 & Story & 0.0 & 1.9 & 46.8 & 51.3 & 0.0 & 0.0 \\
 & Vitruveo & 0.0 & 1.6 & 27.1 & 71.4 & 0.0 & 0.0 \\
 & AirDao & 0.0 & 1.8 & 59.1 & 39.1 & 0.0 & 0.0 \\
 & Etho & 0.0 & 1.6 & 88.7 & 8.1 & 0.0 & 1.6 \\
 & Metis & 0.0 & 2.7 & 45.9 & 51.4 & 0.0 & 0.0 \\
 &Edgeware & 0.0 & 0.0 & 25.0 & 75.0 & 0.0 & 0.0 \\
 & Vana & 0.0 & 1.4 & 36.1 & 62.5 & 0.0 & 0.0 \\
 & Fuse & 0.0 & 0.0 & 95.2 & 4.8 & 0.0 & 0.0 \\

 & Lumoz & 0.0 & 0.0 & 83.3 & 16.7 & 0.0 & 0.0 \\

\midrule

\multirowcell{5}{\rotatebox[origin=c]{90}{\shortstack[c]{Ethereum\\(discv5)}}}

 & Ethereum (Deneb) & 0.0 & 6.4 & 59.7 & 31.8 & 1.3 & 0.7 \\
 & Ethereum (Electra) & 0.3 & 6.2 & 46.5 & 43.3 & 1.7 & 2.0 \\
 & Gnosis (Deneb) & 0.0 & 5.6 & 69.4 & 22.2 & 2.8 & 0.0 \\
 & Gnosis (Electra) & 0.0 & 3.6 & 61.6 & 32.1 & 1.1 & 1.5 \\
 & Ethereum (from discv4) & 0.0 & 10.0 & 32.5 & 55.0 & 2.5 & 0.0 \\

\midrule

\multirowcell{7}{\rotatebox[origin=c]{90}{}}
 & Ripple & 0.2 & 16.6 & 44.3 & 38.6 & 0.2 & 0.0 \\
 & Solana & 2.1 & 0.6 & 29.8 & 65.7 & 0.0 & 1.7 \\
 & Dogecoin & 0.3 & 8.9 & 46.7 & 42.3 & 1.6 & 0.1 \\
 & Cardano & 0.2 & 7.2 & 46.4 & 45.2 & 0.7 & 0.2 \\
 & Avalanche & 0.0 & 2.0 & 33.1 & 64.6 & 0.3 & 0.1 \\
 & Polkadot & 0.4 & 25.5 & 33.1 & 40.1 & 0.2 & 0.7 \\
 & Stellar & 0.0 & 4.9 & 45.9 & 48.6 & 0.5 & 0.0 \\
 \midrule
\multirowcell{3}{\rotatebox[origin=c]{90}{\shortstack[c]{Bitcoin\\ Cash (*)}}}

 & BitcoinCash & 0.0 & 3.3 & 55.6 & 39.3 & 1.8 & 0.0 \\
 & BitcoinSV & 0.0 & 4.5 & 53.7 & 41.8 & 0.0 & 0.0 \\
 & eCash & 0.0 & 19.4 & 52.5 & 27.0 & 1.2 & 0.0 \\
\midrule

\multirowcell{6}{\rotatebox[origin=c]{90}{}}
 & Litecoin & 0.2 & 8.3 & 60.2 & 30.5 & 0.6 & 0.2 \\
 & Near & 0.0 & 0.7 & 60.0 & 39.3 & 0.0 & 0.0 \\
 & Filecoin & 0.0 & 59.2 & 19.3 & 19.3 & 2.0 & 0.2 \\
 & Celestia & 0.7 & 3.5 & 53.1 & 37.1 & 0.7 & 4.9 \\
 & Avail & 1.7 & 23.6 & 35.5 & 33.9 & 3.9 & 1.4 \\
 \midrule
 & Overall & 0.3 & 10.2 & 41.7 & 46.1 & 0.9 & 0.8 \\
\bottomrule
\end{tabular}
}
\caption{Percentage distribution of active blockchain IPs by continent. Values represent the fraction of daily observed nodes across Africa, Asia, Europe, North America, Oceania, and South America, with blockchains in the multi-network discovery protocol broken down.}
\label{tab:continent_dist_all}
\end{table*}

\FloatBarrier

\section{Network Overlap}\label{app:overlap}

\begin{figure}[H]
        \centering
        \includegraphics[scale=1]{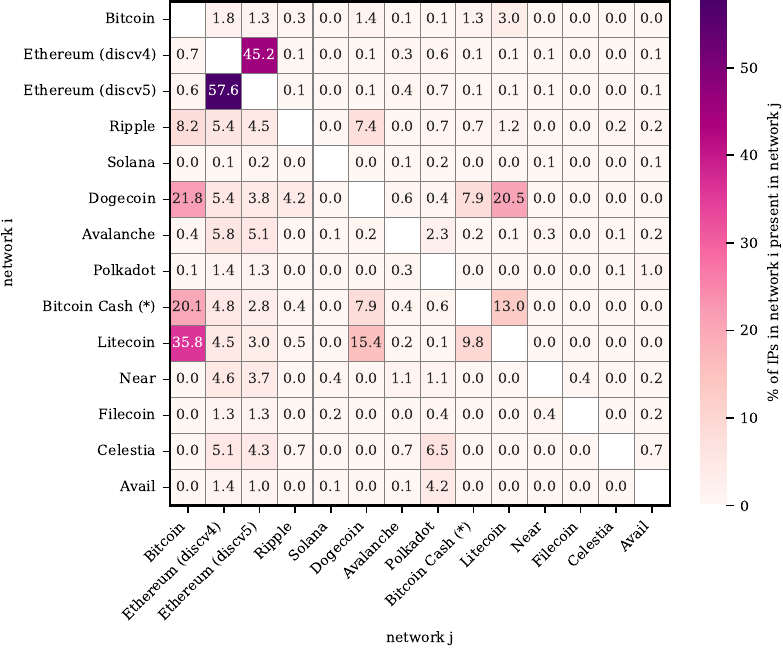}
        \caption{Heatmap of the percentage overlap of active node IPs between blockchain networks on April 10, 2025. Each cell at row $i$, column $j$ shows the percentage of IPs active in network $i$ that are also active in network $j$ on that date. The diagonal is masked, since it trivially represents 100\% self–overlap. Note that Stellar is not included in this figure, as we do not have data for the day.}
        \Description{}
        \label{fig:network_overlap_april10}
    \end{figure}

    We investigate node‐level overlap in Figure~\ref{fig:network_overlap_april10}, which reports the pairwise share of active IPs between every discovery network on April 10, 2025. The strongest overlap (approximately 50\%) occurs between Ethereum (discv4) and Ethereum (discv5), reflecting the requirement to run nodes in both networks when operating an Ethereum node. Recall that Ethereum nodes make up the biggest proportion of the two networks. Further, both Ethereum networks also exhibit a modest overlap (around 5–10\%) with several other networks. This is not unexpected due to the diverse set of blockchains in the networks.

    Among the Bitcoin‐based networks, we observe mutual overlaps of 20–35\% between Bitcoin, Dogecoin, Bitcoin Cash (*) and Litecoin, indicating significant cross‐participation. It is not surprising that operators in these smaller networks would also operate a Bitcoin node. Some networks, such as Ripple and Avalanche, overlap with multiple other chains at the 5–10\% level, whereas Solana shows minimal overlap (< 1\%) with all other crawled networks, suggesting an almost entirely distinct set of peers.  
\end{document}

%% file: figures/overview_table.tex
\begin{table}[t]
\begin{threeparttable}
\resizebox{\textwidth}{!}{
    \begin{tabular}{@{}l|c|c|ccc|c|lc@{}}
    \toprule

 & \textbf{\makecell{Market Cap \\ (12/24)}} & \textbf{Crawler Type} & \textbf{crawl ping} & \textbf{protocol ping}& \textbf{tcp ping}& \textbf{IP Scan} & \textbf{node explorer} & \textbf{scraped} \\
\hline

\textbf{Bitcoin} & \$1937.1B & bitcoin-client & & \checkmark & \checkmark & \checkmark & IPs and stats \cite{blockchair,bitnodes}\tnote{a} & \checkmark\\

\textbf{Ethereum (discv4)} & \$462.8B & DHT & \checkmark && \checkmark &  & IPs and stats \cite{ethernodes}\tnote{b}& \checkmark \\

\textbf{Ethereum (discv5)} &  & DHT & \checkmark & &\checkmark &  & stats \cite{migalabs,ethernodes, probelab} &  \\

\textbf{Ripple}  & \$131.3B & RPC & &\checkmark & \checkmark & & stats\cite{xrpscan_validators,xrpl_peerfinder,bithomp_amendments}&\\

\textbf{Solana} & \$111.2B & API & & & \checkmark & & stats \cite{solanabeach}\tnote{c} & \\

\textbf{Dogecoin} & \$62.8B & bitcoin-client &  & \checkmark & \checkmark & \checkmark & IPs \cite{blockchair} & \checkmark \\

\textbf{Cardano} & \$40.8B & n/a &&&& \checkmark & stats \cite{cardanofoundation} & \\

\textbf{Avalanche} & \$20.7B & RPC &  && \checkmark && stats \cite{avaxnetworkstats}&\\

\textbf{Polkadot}& \$15.6B & DHT & \checkmark & &\checkmark &  & stats \cite{probelab}& \\

\textbf{Stellar} & \$14.0B  & custom & \checkmark &  & \checkmark & & IP and stats \cite{obsrvrRadar} & \checkmark \\

\textbf{Bitcoin Cash (*)}& \$11.9B & bitcoin-client & & \checkmark & \checkmark & \checkmark & IPs \cite{blockchair} & \checkmark \\

 \textbf{Litecoin} & \$10.0B  & bitcoin-client &  & \checkmark & \checkmark &  & stats \cite{blockchair} & \checkmark \\

\textbf{Near} & \$9.4B & RPC & \checkmark & \checkmark & \checkmark & & & \\

\textbf{Filecoin}  & \$3.2B & DHT & \checkmark && \checkmark & &  stats \cite{probelab} & \\

\textbf{Celestia} & \$2.5B &  DHT & \checkmark && \checkmark & & stats \cite{probelab} & \\

\textbf{Avail} & \$0.19B &   DHT & \checkmark && \checkmark & & stats \cite{probelab} & \\

\bottomrule
\end{tabular}}
\begin{tablenotes}
\scriptsize 
\item[a] We scrape both explorers; Blockchair only provides IPv4 nodes, while Bitnodes includes IPv6 and Tor.
\item[b] \texttt{ethernodes.org} historically provided discovered IPs, but since March 2025 shows only aggregate counts.
\item[c] Their API used to expose IPs for non-validator nodes; as of mid-May 2025 it no longer does.
\end{tablenotes}
\vspace{4pt}
\caption{Networks we measure individually, ordered by their market capitalization at the start of our measurements (December 24, 2024). Cardano is the one network in the list for which we do not run daily measurements, but instead measure via two IP scans. All others we run daily crawler and connectivity checks.
}
    \label{tab:summary-protocols}
\vspace{-20pt}
\end{threeparttable}
\end{table}